\documentclass{article}

\usepackage{type1cm}        

\usepackage{makeidx}
\usepackage{graphicx}
\usepackage{multicol}
\usepackage[bottom]{footmisc}
\usepackage{newtxtext}
\usepackage[varvw]{newtxmath}
\usepackage{tikz}
\usepackage{cite}
\usepackage{subcaption}
\usepackage{algorithm}
\usepackage{algpseudocode}
\usepackage{listings}
\usepackage{hyperref}
\usepackage{authblk}

\definecolor{ownblue}{RGB}{12,123,220}
\definecolor{pythonblue}{RGB}{0,0,255}

\definecolor{ownorange}{RGB}{255,194,10}
\definecolor{pythonorange}{RGB}{255,184,54}

\definecolor{ownbrown}{RGB}{153,79,0}

\definecolor{owngreen}{RGB}{26,255,26}
\definecolor{pythongreen}{RGB}{0,128,0}

\definecolor{pythongray}{RGB}{128,128,128}

\definecolor{darkgreen}{RGB}{30,140,20}

\makeindex

\begin{document}

\title{Tuning of Vectorization Parameters for Molecular Dynamics Simulations in AutoPas}

\author[1]{Luis Gall}
\author[1]{Samuel James Newcome}
\author[1]{Fabio Alexander Gratl}
\author[1]{Markus Mühlhäußer}
\author[1]{Manish Kumar Mishra}
\author[1]{Hans-Joachim Bungartz}

\affil[1]{Chair of Scientific Computing in Computer Science, Department of Computer Science,  Technical University of Munich, Boltzmannstr. 3, 85748, Garching bei Muenchen, Germany; luis.gall@tum.de}

\date{}

\maketitle

\begin{abstract}
    Molecular Dynamics simulations can help scientists to gather valuable insights for physical processes on an atomic scale.
    This work explores various techniques for SIMD vectorization to improve the pairwise force calculation between molecules in the scope of the particle simulation library AutoPas.
    The focus lies on the order in which particle values are loaded into vector registers to achieve the most optimal performance regarding execution time or energy consumption.
    \newline
    As previous work indicates that the optimal MD algorithm can change during runtime, this paper investigates simulation-specific parameters like particle density and the impact of the neighbor identification algorithms, which distinguishes this work from related projects.
    Furthermore, AutoPas' dynamic tuning mechanism is extended to choose the optimal vectorization order during runtime.
    \newline
    The benchmarks show that considering different particle interaction orders during runtime can lead to a considerable performance improvement for the force calculation compared to AutoPas' previous approach.
\end{abstract}

\section{Introduction}
\label{chap:introduction}

Molecular Dynamics (MD) simulations provide wide possibilities for studying physical processes on a molecular scale.
Applications of MD models include investigating the crack propagation behavior of nanocrystalline metals \cite{Crack-propagation} and deriving viscosity correlations of fluids \cite{application_fluid_2005}.

Frequently, the calculation of the pairwise intermolecular forces is the computationally dominant part of MD software \cite{Gratl_Autopas_2022}.
In addition to several algorithmic improvements that eliminate unnecessary computations, modern high-performance computing (HPC) techniques have been successfully applied to enable bigger simulations \cite{trillion_2019}.
These optimizations include intelligent memory and cache utilization, usage of processors' SIMD (Single Instruction, Multiple Data) units, multithreading, and GPU-based acceleration \cite{gromacs_overview_2020}.

However, developers face challenges maintaining their code with a growing diversity of modern compute clusters, such as GPU accelerators or various CPU architectures (Intel and AMD vs. ARM) \cite{arm_intel_2021}, which introduces the importance of performance portability.
Although the meaning of this term is debated \cite{Pennycook_performance_portability_2019}, one can refer to it as the ability to write code portable to multiple architectures without sacrificing performance.

In addition to abstracting architecture-specific implementations, non-HPC experts might struggle to tune their code to every possible simulation scenario because no algorithm is performing best in every application and on every machine \cite{Gratl_Autopas_2022}.
For vectorization, some code might only perform well on architectures that efficiently support sub-register accesses \cite{gromacs_overview_2020}.
To tackle these sorts of challenges, the open-source MD library AutoPas\footnote{\url{https://github.com/AutoPas/AutoPas}} is developed, which offers a wide range of MD algorithms and optimizations.
Its tuning mechanisms aim to choose the optimal configuration for the given simulation environment at runtime, including architecture-specific optimizations. \cite{Gratl_Autopas_2022}

Apart from abstracting CPU vendor-specific intrinsics, vectorization in MD introduces the question of which order particle values should be loaded into the SIMD registers \cite{ls1mardyn_2020}.
When looping over the pairs of particle lists in a nested fashion, the interaction order determines the number of distinct particles loaded from each list.
Other popular particle simulators, including GROMACS\cite{Gromacs_simd_2015} and LAMMPS \cite{lammps_tersoff_2016}, mainly tune their vectorization orders to one specific neighbor identification algorithm.
This paper considers runtime-specific influences, such as changing particle densities and the impact of different neighbor identification algorithms, and investigates whether the optimal vectorization order can change during runtime.
Finally, AutoPas' tuning mechanism is extended to choose the fastest-time-to-solution or smallest energy consumption particle interaction order at runtime.

This paper first clarifies the theoretical and technical background in \autoref{chap:background}.
Next, an overview of related projects is provided in \autoref{chap:related_work} followed by a discussion of the SIMD implementation in \autoref{chap:implementation}.
Afterward, \autoref{chap:results} discusses the benchmark results, and \autoref{chap:future} includes an outlook on possible future work.
Finally, a conclusion is drawn in \autoref{chap:conclusion}.

\section{Theoretical and Technical Background}
\label{chap:background}

This section highlights the theoretical background of Molecular Dynamics, the technical concept of vectorization, and how it is employed in MD.
Furthermore, the relevant background of the AutoPas ecosystem is highlighted.

\subsection{Molecular Dynamics}

Molecular Dynamics simulations form a subgroup of physics-based computer simulations.
Instead of considering flow or heat equations, MD computes forces acting on molecules and translates these forces to movement based on Verlet integration for Newton's Laws of Motion \cite{Gratl_Autopas_2022}.
Usually, the force acting on a single particle $i$ is obtained by the sum of forces exerted by all other particles on $i$ \cite{Santamaria_MD_2023}:

\begin{equation}
    F_i = \sum_{j \in \text{particles}} F_{ij}
    \label{eq:force_sum}
\end{equation}

Consequently, computing the forces between these particles is computationally intensive because of the complexity of $\mathcal{O}(N^2)$ with $N$ as the number of particles in the system \cite{Gratl_Autopas_2022}.

A first optimization can be Newton's third law, stating that $F_{ij} = -F_{ji}$ holds for particles $i$ and $j$.
With this help, the number of force calculations can be reduced by a factor of two.
However, the quadratic complexity persists. \cite{Santamaria_MD_2023}

As the value of the force converges quickly towards zero with a growing distance between particles, only the calculations between particles within a certain distance are relevant.
This distance is often referred to as the cutoff, and a key challenge in molecular dynamics simulation is the identification of particles inside the cutoff. \cite{Gratl_Autopas_2022}

\subsubsection{Neighbor Identification Algorithms}
\label{sec:neighbor_algorithms}
\begin{figure}[h]
    \centering
    \begin{subfigure}[b]{0.3\textwidth}
        \centering
        \begin{tikzpicture}[scale=0.675]
            \fill[ownorange!20] (1,1) rectangle (4,4);
            \fill[ownblue!20] (2.5,2.5) circle (1.05);
            \draw (0,0) grid (5,5);
            \pgfmathsetseed{5}
            \foreach \x in {0.5,1.5,2.5,3.5,4.3} {
                \foreach \y in {0.5,4.3} {

                    \coordinate (A) at (\x,\y);
                    \coordinate (B) at (\x+0.1+rand/7,\y+0.3+rand/7);
                    \coordinate (C) at (\x+0.3+rand/7,\y+0.2+rand/7);
                    \coordinate (D) at (\x+0.4+rand/7,\y-0.1+rand/7);

                    \fill (A) circle (1.25pt);
                    \fill (B) circle (1.25pt);
                    \fill (C) circle (1.25pt);
                    \fill (D) circle (1.25pt);
                }
            }
            \foreach \x in {0.3,4.3} {
                \foreach \y in {1.5,2.5,3.5} {

                    \coordinate (A) at (\x,\y);
                    \coordinate (B) at (\x+0.1+rand/7,\y+0.3+rand/7);
                    \coordinate (C) at (\x+0.3+rand/7,\y+0.2+rand/7);
                    \coordinate (D) at (\x+0.4+rand/7,\y-0.1+rand/7);

                    \fill (A) circle (1.25pt);
                    \fill (B) circle (1.25pt);
                    \fill (C) circle (1.25pt);
                    \fill (D) circle (1.25pt);
                }
            }
            \foreach \x in {2.35} {
                \foreach \y in {1.5,3.5} {

                    \coordinate (A) at (\x,\y);
                    \coordinate (B) at (\x+0.1+rand/7,\y+0.3+rand/7);
                    \coordinate (C) at (\x+0.3+rand/7,\y+0.2+rand/7);
                    \coordinate (D) at (\x+0.4+rand/7,\y-0.1+rand/7);

                    \fill [ownorange!90] (A) circle (2pt);
                    \draw [gray!85] (A) circle (2pt);
                    \fill [ownorange!90] (B) circle (2pt);
                    \draw [gray!85] (B) circle (2pt);
                    \fill [ownorange!90] (C) circle (2pt);
                    \draw [gray!85] (C) circle (2pt);
                    \fill [ownorange!90] (D) circle (2pt);                    
                    \draw [gray!85] (D) circle (2pt);
                }
            }
            \foreach \x in {1.35} {
                \foreach \y in {1.5,2.5,3.5} {

                    \coordinate (A) at (\x,\y);
                    \coordinate (B) at (\x+0.1+rand/7,\y+0.3+rand/7);
                    \coordinate (C) at (\x+0.3+rand/7,\y+0.2+rand/7);
                    \coordinate (D) at (\x+0.4+rand/7,\y-0.1+rand/7);

                    \fill [ownorange!90] (A) circle (2pt);
                    \draw [gray!85] (A) circle (2pt);
                    \fill [ownorange!90] (B) circle (2pt);
                    \draw [gray!85] (B) circle (2pt);
                    \fill [ownorange!90] (C) circle (2pt);
                    \draw [gray!85] (C) circle (2pt);
                    \fill [ownorange!90] (D) circle (2pt);                    
                    \draw [gray!85] (D) circle (2pt);
                }
            }
            \foreach \x in {3.35} {
                \foreach \y in {1.5,2.5,3.5} {

                    \coordinate (A) at (\x,\y);
                    \coordinate (B) at (\x+0.1+rand/7,\y+0.3+rand/7);
                    \coordinate (C) at (\x+0.3+rand/7,\y+0.2+rand/7);
                    \coordinate (D) at (\x+0.4+rand/7,\y-0.1+rand/7);

                    \fill [ownorange!90] (A) circle (2pt);
                    \draw [gray!85] (A) circle (2pt);
                    \fill [ownorange!90] (B) circle (2pt);
                    \draw [gray!85] (B) circle (2pt);
                    \fill [ownorange!90] (C) circle (2pt);
                    \draw [gray!85] (C) circle (2pt);
                    \fill [ownorange!90] (D) circle (2pt);                    
                    \draw [gray!85] (D) circle (2pt);
                }
            }
            \foreach \x in {2.35} {
                \foreach \y in {2.5} {

                    \coordinate (A) at (\x,\y);
                    \coordinate (B) at (\x+0.1+rand/7,\y+0.3+rand/7);
                    \coordinate (C) at (\x+0.3+rand/7,\y+0.2+rand/7);
                    \coordinate (D) at (\x+0.4+rand/7,\y-0.1+rand/7);

                    \fill [ownblue!90] (A) circle (2pt);
                    \draw [gray!85] (A) circle (2pt);
                    \fill [ownblue!90] (B) circle (2pt);
                    \draw [gray!85] (B) circle (2pt);
                    \fill [ownblue!90] (C) circle (2pt);
                    \draw [gray!85] (C) circle (2pt);
                    \fill [ownblue!90] (D) circle (2pt);                    
                    \draw [gray!85] (D) circle (2pt);
                }
            }
        \end{tikzpicture}
        
        \subcaption{Linked Cells, illustration inspired by \cite{Multisite_2023}}
        \label{fig:linked_cells}
    \end{subfigure}
    \hfill
    \begin{subfigure}[b]{0.3\textwidth}
        \centering
        \begin{tikzpicture}[scale=0.675]
            \fill[ownorange!20] (2.5,2.5) circle (1.275);
            \fill[ownblue!20] (2.5,2.5) circle (1.05);
            \draw (0,0) rectangle (5,5);
            \pgfmathsetseed{5}
            \foreach \x in {0.5,1.5,2.5,3.5,4.3} {
                \foreach \y in {0.5,4.3} {

                    \coordinate (A) at (\x,\y);
                    \coordinate (B) at (\x+0.1+rand/7,\y+0.3+rand/7);
                    \coordinate (C) at (\x+0.3+rand/7,\y+0.2+rand/7);
                    \coordinate (D) at (\x+0.4+rand/7,\y-0.1+rand/7);

                    \fill (A) circle (1.25pt);
                    \fill (B) circle (1.25pt);
                    \fill (C) circle (1.25pt);
                    \fill (D) circle (1.25pt);
                }
            }
            \foreach \x in {0.3,4.3} {
                \foreach \y in {1.5,2.5,3.5} {

                    \coordinate (A) at (\x,\y);
                    \coordinate (B) at (\x+0.1+rand/7,\y+0.3+rand/7);
                    \coordinate (C) at (\x+0.3+rand/7,\y+0.2+rand/7);
                    \coordinate (D) at (\x+0.4+rand/7,\y-0.1+rand/7);

                    \fill (A) circle (1.25pt);
                    \fill (B) circle (1.25pt);
                    \fill (C) circle (1.25pt);
                    \fill (D) circle (1.25pt);
                }
            }
            \foreach \x in {2.35} {
                \foreach \y in {1.5,3.5} {

                    \coordinate (A) at (\x,\y);
                    \coordinate (B) at (\x+0.1+rand/7,\y+0.3+rand/7);
                    \coordinate (C) at (\x+0.3+rand/7,\y+0.2+rand/7);
                    \coordinate (D) at (\x+0.4+rand/7,\y-0.1+rand/7);

                    \fill [ownorange!90] (A) circle (2pt);
                    \draw [gray!85] (A) circle (2pt);
                    \fill [ownorange!90] (B) circle (2pt);
                    \draw [gray!85] (B) circle (2pt);
                    \fill [ownorange!90] (C) circle (2pt);
                    \draw [gray!85] (C) circle (2pt);
                    \fill [ownorange!90] (D) circle (2pt);                    
                    \draw [gray!85] (D) circle (2pt);
                }
            }
            \foreach \x in {1.35} {
                \foreach \y in {1.5} {

                    \coordinate (A) at (\x,\y);
                    \coordinate (B) at (\x+0.1+rand/7,\y+0.3+rand/7);
                    \coordinate (C) at (\x+0.3+rand/7,\y+0.2+rand/7);
                    \coordinate (D) at (\x+0.4+rand/7,\y-0.1+rand/7);

                    \fill (A) circle (1.25pt);
                    \fill [ownorange!90] (B) circle (2pt);
                    \draw [gray!85] (B) circle (2pt);
                    \fill [ownorange!90] (C) circle (2pt);
                    \draw [gray!85] (C) circle (2pt);
                    \fill (D) circle (1.25pt);
                }
                \foreach \y in {2.5} {

                    \coordinate (A) at (\x,\y);
                    \coordinate (B) at (\x+0.1+rand/7,\y+0.3+rand/7);
                    \coordinate (C) at (\x+0.3+rand/7,\y+0.2+rand/7);
                    \coordinate (D) at (\x+0.4+rand/7,\y-0.1+rand/7);

                    \fill [ownorange!90] (A) circle (2pt);
                    \draw [gray!85] (A) circle (2pt);
                    \fill [ownorange!90] (B) circle (2pt);
                    \draw [gray!85] (B) circle (2pt);
                    \fill [ownorange!90] (C) circle (2pt);
                    \draw [gray!85] (C) circle (2pt);
                    \fill [ownorange!90] (D) circle (2pt);                    
                    \draw [gray!85] (D) circle (2pt);
                }
                \foreach \y in {3.5} {

                    \coordinate (A) at (\x,\y);
                    \coordinate (B) at (\x+0.1+rand/7,\y+0.3+rand/7);
                    \coordinate (C) at (\x+0.3+rand/7,\y+0.2+rand/7);
                    \coordinate (D) at (\x+0.4+rand/7,\y-0.1+rand/7);

                    \fill (A) circle (1.25pt);
                    \fill (B) circle (1.25pt);
                    \fill (C) circle (1.25pt);
                    \fill [ownorange!90] (D) circle (2pt);                    
                    \draw [gray!85] (D) circle (2pt);
                }
            }
            \foreach \x in {3.35} {
                \foreach \y in {1.5} {

                    \coordinate (A) at (\x,\y);
                    \coordinate (B) at (\x+0.1+rand/7,\y+0.3+rand/7);
                    \coordinate (C) at (\x+0.3+rand/7,\y+0.2+rand/7);
                    \coordinate (D) at (\x+0.4+rand/7,\y-0.1+rand/7);

                    \fill [ownorange!90] (A) circle (2pt);
                    \draw [gray!85] (A) circle (2pt);
                    \fill [ownorange!90] (B) circle (2pt);
                    \draw [gray!85] (B) circle (2pt);
                    \fill [ownorange!90] (C) circle (2pt);
                    \draw [gray!85] (C) circle (2pt);
                    \fill (D) circle (1.25pt);
                }
                \foreach \y in {2.5} {

                    \coordinate (A) at (\x,\y);
                    \coordinate (B) at (\x+0.1+rand/7,\y+0.3+rand/7);
                    \coordinate (C) at (\x+0.3+rand/7,\y+0.2+rand/7);
                    \coordinate (D) at (\x+0.4+rand/7,\y-0.1+rand/7);

                    \fill [ownorange!90] (A) circle (2pt);
                    \draw [gray!85] (A) circle (2pt);
                    \fill [ownorange!90] (B) circle (2pt);
                    \draw [gray!85] (B) circle (2pt);
                    \fill [ownorange!90] (C) circle (2pt);
                    \draw [gray!85] (C) circle (2pt);
                    \fill (D) circle (1.25pt);
                }
                \foreach \y in {3.5} {

                    \coordinate (A) at (\x,\y);
                    \coordinate (B) at (\x+0.1+rand/7,\y+0.3+rand/7);
                    \coordinate (C) at (\x+0.3+rand/7,\y+0.2+rand/7);
                    \coordinate (D) at (\x+0.4+rand/7,\y-0.1+rand/7);

                    \fill (A) circle (1.25pt);
                    \fill (B) circle (1.25pt);
                    \fill (C) circle (1.25pt);
                    \fill (D) circle (1.25pt);
                }
            }
            \foreach \x in {2.35} {
                \foreach \y in {2.5} {
                    \coordinate (A) at (2.5,2.5);
                    \coordinate (B) at (\x+0.3+rand/7,\y+0.5+rand/7);
                    \coordinate (C) at (\x+0.6+rand/7,\y+0.3+rand/7);
                    \coordinate (D) at (\x+0.7+rand/7,\y-0.3+rand/7);

                    \fill [ownblue!90] (A) circle (2pt);
                    \draw [gray!85] (A) circle (2pt);
                    \fill [ownorange!90] (B) circle (2pt);
                    \draw [gray!85] (B) circle (2pt);
                    \fill [ownorange!90] (C) circle (2pt);
                    \draw [gray!85] (C) circle (2pt);
                    \fill [ownorange!90] (D) circle (2pt);                    
                    \draw [gray!85] (D) circle (2pt);
                }
            }
        \end{tikzpicture}
        
        \subcaption{Verlet Lists, illustration inspired by \cite{Multisite_2023}}
        \label{fig:verlet_lists}
    \end{subfigure}
    \hfill
    \begin{subfigure}[b]{0.3\textwidth}
        \centering
        \begin{tikzpicture}[scale=0.675]
            
            \draw (0,0) rectangle (5,5);
            
            \fill [ownorange!25] (2.5,2.5) circle (1.275);
            \fill [ownblue!20] (2.5,2.5) circle (1.05);
            \pgfmathsetseed{5}
            \foreach \x in {0.5,1.5,2.5,3.5,4.3} {
                \foreach \y in {0.5,4.3} {

                    \coordinate (A) at (\x,\y);
                    \coordinate (B) at (\x+0.1+rand/7,\y+0.3+rand/7);
                    \coordinate (C) at (\x+0.3+rand/7,\y+0.2+rand/7);
                    \coordinate (D) at (\x+0.4+rand/7,\y-0.1+rand/7);

                    \draw [dashed, gray!90] (A) -- (B);
                    \draw [dashed, gray!90] (B) -- (C);
                    \draw [dashed, gray!90] (C) -- (D);
                    \draw [dashed, gray!90] (D) -- (A);

                    \fill (A) circle (1.25pt);
                    \fill (B) circle (1.25pt);
                    \fill (C) circle (1.25pt);
                    \fill (D) circle (1.25pt);
                }
            }
            \foreach \x in {0.5,4.3} {
                \foreach \y in {1.5,2.5,3.5} {

                    \coordinate (A) at (\x,\y);
                    \coordinate (B) at (\x+0.1+rand/7,\y+0.3+rand/7);
                    \coordinate (C) at (\x+0.3+rand/7,\y+0.2+rand/7);
                    \coordinate (D) at (\x+0.4+rand/7,\y-0.1+rand/7);

                    \draw [dashed, gray!90] (A) -- (B);
                    \draw [dashed, gray!90] (B) -- (C);
                    \draw [dashed, gray!90] (C) -- (D);
                    \draw [dashed, gray!90] (D) -- (A);

                    \fill (A) circle (1.25pt);
                    \fill (B) circle (1.25pt);
                    \fill (C) circle (1.25pt);
                    \fill (D) circle (1.25pt);
                }
            }
            \foreach \x in {2.5} {
                \foreach \y in {1.5,3.5} {

                    \coordinate (A) at (\x,\y);
                    \coordinate (B) at (\x+0.1+rand/7,\y+0.3+rand/7);
                    \coordinate (C) at (\x+0.3+rand/7,\y+0.2+rand/7);
                    \coordinate (D) at (\x+0.4+rand/7,\y-0.1+rand/7);

                    \draw [dashed, gray!90] (A) -- (B);
                    \draw [dashed, gray!90] (B) -- (C);
                    \draw [dashed, gray!90] (C) -- (D);
                    \draw [dashed, gray!90] (D) -- (A);

                    \fill [ownorange!90] (A) circle (2pt);
                    \draw [gray!85] (A) circle (2pt);
                    \fill [ownorange!90] (B) circle (2pt);
                    \draw [gray!85] (B) circle (2pt);
                    \fill [ownorange!90] (C) circle (2pt);
                    \draw [gray!85] (C) circle (2pt);
                    \fill [ownorange!90] (D) circle (2pt);
                    \draw [gray!85] (D) circle (2pt);
                }
            }
            \foreach \x in {1.5} {
                \foreach \y in {1.5,2.5,3.5} {

                    \coordinate (A) at (\x,\y);
                    \coordinate (B) at (\x+0.1+rand/7,\y+0.3+rand/7);
                    \coordinate (C) at (\x+0.3+rand/7,\y+0.2+rand/7);
                    \coordinate (D) at (\x+0.4+rand/7,\y-0.1+rand/7);

                    \draw [dashed, gray!90] (A) -- (B);
                    \draw [dashed, gray!90] (B) -- (C);
                    \draw [dashed, gray!90] (C) -- (D);
                    \draw [dashed, gray!90] (D) -- (A);

                    \fill [ownorange!90] (A) circle (2pt);
                    \draw [gray!85] (A) circle (2pt);
                    \fill [ownorange!90] (B) circle (2pt);
                    \draw [gray!85] (B) circle (2pt);
                    \fill [ownorange!90] (C) circle (2pt);
                    \draw [gray!85] (C) circle (2pt);
                    \fill [ownorange!90] (D) circle (2pt);                    
                    \draw [gray!85] (D) circle (2pt);
                }
            }

            \foreach \x in {3.5} {
                \foreach \y in {1.5} {

                    \coordinate (A) at (\x,\y);
                    \coordinate (B) at (\x+0.1+rand/7,\y+0.3+rand/7);
                    \coordinate (C) at (\x+0.3+rand/7,\y+0.2+rand/7);
                    \coordinate (D) at (\x+0.4+rand/7,\y-0.1+rand/7);

                    \draw [dashed, gray!90] (A) -- (B);
                    \draw [dashed, gray!90] (B) -- (C);
                    \draw [dashed, gray!90] (C) -- (D);
                    \draw [dashed, gray!90] (D) -- (A);

                    \fill (A) circle (1.25pt);
                    \draw [gray!85] (A) circle (1.25pt);
                    \fill (B) circle (1.25pt);
                    \draw [gray!85] (B) circle (1.25pt);
                    \fill (C) circle (1.25pt);
                    \draw [gray!85] (C) circle (1.25pt);
                    \fill (D) circle (1.25pt);
                    \draw [gray!85] (D) circle (1.25pt);
                }
            }
            
            \foreach \x in {3.5} {
                \foreach \y in {2.5} {

                    \coordinate (A) at (\x,\y);
                    \coordinate (B) at (\x+0.1+rand/7,\y+0.3+rand/7);
                    \coordinate (C) at (\x+0.3+rand/7,\y+0.2+rand/7);
                    \coordinate (D) at (\x+0.4+rand/7,\y-0.1+rand/7);

                    \draw [dashed, gray!90] (A) -- (B);
                    \draw [dashed, gray!90] (B) -- (C);
                    \draw [dashed, gray!90] (C) -- (D);
                    \draw [dashed, gray!90] (D) -- (A);

                    \fill [ownorange!90] (A) circle (2pt);
                    \draw [gray!85] (A) circle (2pt);
                    \fill [ownorange!90] (B) circle (2pt);
                    \draw [gray!85] (B) circle (2pt);
                    \fill [ownorange!90] (C) circle (2pt);
                    \draw [gray!85] (C) circle (2pt);
                    \fill [ownorange!90] (D) circle (2pt);
                    \draw [gray!85] (D) circle (2pt);
                }
            }
            \foreach \x in {3.5} {
                \foreach \y in {3.5} {

                    \coordinate (A) at (\x,\y);
                    \coordinate (B) at (\x+0.1+rand/7,\y+0.3+rand/7);
                    \coordinate (C) at (\x+0.3+rand/7,\y+0.2+rand/7);
                    \coordinate (D) at (\x+0.4+rand/7,\y-0.1+rand/7);

                    \draw [dashed, gray!90] (A) -- (B);
                    \draw [dashed, gray!90] (B) -- (C);
                    \draw [dashed, gray!90] (C) -- (D);
                    \draw [dashed, gray!90] (D) -- (A);

                    \fill (A) circle (1.25pt);
                    \draw [gray!85] (A) circle (1.25pt);
                    \fill (B) circle (1.25pt);
                    \draw [gray!85] (B) circle (1.25pt);
                    \fill (C) circle (1.25pt);
                    \draw [gray!85] (C) circle (1.25pt);
                    \fill (D) circle (1.25pt);
                    \draw [gray!85] (D) circle (1.25pt);
                }
            }
            \foreach \x in {2.5} {
                \foreach \y in {2.5} {

                    \coordinate (A) at (\x,\y);
                    \coordinate (B) at (\x+0.1+rand/7,\y+0.3+rand/7);
                    \coordinate (C) at (\x+0.3+rand/7,\y+0.2+rand/7);
                    \coordinate (D) at (\x+0.4+rand/7,\y-0.1+rand/7);

                    \draw [dashed, gray!90] (A) -- (B);
                    \draw [dashed, gray!90] (B) -- (C);
                    \draw [dashed, gray!90] (C) -- (D);
                    \draw [dashed, gray!90] (D) -- (A);

                    \fill [ownblue!90] (A) circle (2pt);
                    \draw [gray!85] (A) circle (2pt);
                    \fill [ownblue!90] (B) circle (2pt);
                    \draw [gray!85] (B) circle (2pt);
                    \fill [ownblue!90] (C) circle (2pt);
                    \draw [gray!85] (C) circle (2pt);
                    \fill [ownblue!90] (D) circle (2pt);
                    \draw [gray!85] (D) circle (2pt);
                }
            }
        \end{tikzpicture}
        \subcaption{Verlet Cluster Lists, illustration inspired by \cite{Pall_vcl_2013}}
        \label{fig:cluster_lists}
    \end{subfigure}
    \caption{Chosen neighbor identification algorithms are visualized. The blue circle highlights the cutoff distance. The blue and orange area highlights the search space in which neighboring particles, also in orange, are considered.}
    \label{fig:containers}
\end{figure}
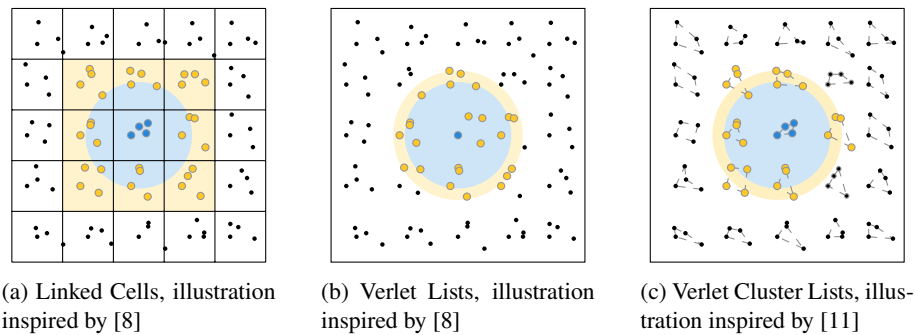

Over the last decades, multiple algorithms have emerged to identify the relevant neighboring particles for force calculation.
The Linked Cells algorithm visualized in \autoref{fig:linked_cells} follows the idea of a spatial decomposition of the simulation domain and divides it into cells of equal size.
Only particles in the same cell (blue particles) and neighboring cells (orange particles) are considered during the force computation.
As the particles nearby in space also end up close in memory, the cache is expected to perform well.
Additionally, as particles are processed sequentially, this algorithm allows for good vectorization. \cite{Gratl_Autopas_2022}

Another idea was proposed by Loup Verlet \cite{verlet_1967} and stores neighborship relations for each particle.
These neighborship buffers, also called neighbor lists or Verlet Lists, have to be updated every few timesteps, as the neighbor status can change.
Therefore, one often includes particles inside the cutoff radius and a skin around the cutoff in the lists.
To simplify the generation of these neighbor lists, the Linked Cells container is often used.\cite{Gratl_Autopas_2022}

A visual explanation of the area where particles are inserted into the lists is provided in \autoref{fig:verlet_lists} (cutoff in blue and skin in orange).
However, this algorithm in its pure form is ill-suited for vectorization as the particles need to be loaded from arbitrary memory locations as the neighbor lists only store references \cite{Gratl_Autopas_2022}.
As vectorization does not allow for different interaction orders in its original form, the algorithm will not be considered further in this paper.

On the other hand, the Verlet Cluster Lists idea suggested by Páll and Hess \cite{Pall_vcl_2013} is specially designed for SIMD parallelization.
It groups particles into fixed-sized clusters, usually aligned to the SIMD width.
Every few iterations, the neighbor lists are constructed to hold full clusters of particles.
The structure of the cluster lists is visually explained in \autoref{fig:cluster_lists} with the blue base cluster and the orange neighboring clusters.

\subsection{Vectorization}
With the concept of SIMD, modern processors can work on multiple data points with one instruction.
CPU vendors typically provide developers with special functions, often called intrinsics, to access these special registers from higher-level programming languages like C++.

In the MD force calculation, two nested loops iterate over particles to compute pairwise interactions following \autoref{eq:force_sum}.
The general vectorization idea is to load different sets of particles in two registers and perform operations to compute multiple distinct particle interactions by one instruction.
Other MD simulators like GROMACS \cite{Gromacs_simd_2015} and LAMMPS \cite{lammps_tersoff_2016} have already examined some of the challenges of vectorization in addition to its potential.

\subsection{AutoPas}

In this work, some vectorization parameters are investigated in the scope of the open-source particle simulation library AutoPas.
It offers various algorithms and optimizations, including implementations for Linked Cells, Verlet Lists, and Verlet Cluster Lists, and aims to choose the best-performing configuration for the current simulation environment at runtime.
\cite{Gratl_Autopas_2022}

In AutoPas, particles are stored in so-called containers, which are concrete implementations of the neighbor identification algorithms (\autoref{fig:containers}).
Every container provides traversal mechanisms that define the order in which the simulation domain is processed, which is important to synchronize particle access in multithreaded simulations.
In the case of the Linked Cells algorithm, for example, the traversal defines the order in which the cells interact.

\begin{figure}[h]
    \centering
    \begin{subfigure}[b]{0.3\textwidth}
        \centering
        \begin{tikzpicture}[scale=0.85]
            \fill[ownorange!35] (1,1) rectangle (4,4);
            \fill[ownblue!25] (2,2) rectangle (3,3);
            \draw (1,1) grid (4,4);
            \foreach \y in {2,3,4} {
                \foreach \x in {2,3,4} {
                    \draw[gray] (\x-0.5,\y-0.5) node {\pgfmathparse{int(\x-1+(\y-2)*3-1)} \pgfmathresult};
                }
            }
            \draw[thick,-stealth] (2.3,2.5) -- (1.7,2.5);
            \draw[thick,-stealth] (2.7,2.5) -- (3.3,2.5);
            \draw[thick,-stealth] (2.5,2.7) -- (2.5,3.3);
            \draw[thick,-stealth] (2.5,2.3) -- (2.5,1.7);
            \draw[thick,-stealth] (2.3,2.3) -- (1.7,1.7);
            \draw[thick,-stealth] (2.7,2.7) -- (3.3,3.3);
            \draw[thick,-stealth] (2.3,2.7) -- (1.7,3.3);
            \draw[thick,-stealth] (2.7,2.3) -- (3.3,1.8);
        \end{tikzpicture}
        \caption{C01 base step, illustration inspired by \cite{Gratl_Autopas_2022}}
        \label{fig:c01 base step}
    \end{subfigure}
    \hfill
    \begin{subfigure}[b]{0.3\textwidth}
        \centering
        \begin{tikzpicture}[scale=0.85]

            \fill[ownorange!35] (1,3) rectangle (4,4);
            \fill[ownorange!35] (3,2) rectangle (4,3);
            \fill[ownblue!25] (2,2) rectangle (3,3);
            \draw (1,1) grid (4,4);
            \foreach \y in {2,3,4} {
                \foreach \x in {2,3,4} {
                    \draw[gray] (\x-0.5,\y-0.5) node {\pgfmathparse{int(\x-1+(\y-2)*3-1)} \pgfmathresult};
                }
            }
            \draw[thick,stealth-stealth] (2.7,2.5) -- (3.3,2.5);
            \draw[thick,stealth-stealth] (2.5,2.7) -- (2.5,3.3);
            \draw[thick,stealth-stealth] (2.7,2.7) -- (3.3,3.3);
            \draw[thick,stealth-stealth] (2.3,2.7) -- (1.7,3.3);
        \end{tikzpicture}
        \caption{C18 base step, illustration inspired by \cite{Gratl_Autopas_2022}}
        \label{fig:c18 base step}
    \end{subfigure}
    \hfill
    \begin{subfigure}[b]{0.3\textwidth}
        \centering
        \begin{tikzpicture}[scale=0.85]

            \fill[ownorange!35] (2,2) rectangle (4,4);
            \fill[ownblue!25] (2,2) rectangle (3,3);
            \draw (1,1) grid (4,4);
            \foreach \y in {2,3,4} {
                \foreach \x in {2,3,4} {
                    \draw[gray] (\x-0.5,\y-0.5) node {\pgfmathparse{int(\x-1+(\y-2)*3-1)} \pgfmathresult};
                }
            }
            \draw[thick,stealth-stealth] (2.7,2.5) -- (3.3,2.5);
            \draw[thick,stealth-stealth] (2.5,2.7) -- (2.5,3.3);
            \draw[thick,stealth-stealth] (2.7,2.7) -- (3.3,3.3);
            \draw[thick,stealth-stealth] (2.7,3.3) -- (3.3,2.7);
        \end{tikzpicture}
        \caption{C08 base step, illustration inspired by \cite{Gratl_Autopas_2022}}
        \label{fig:c08 base step}
    \end{subfigure}
    \caption{The orange cell is the currently processed cell by the domain traversal. The blue area and the black arrows indicate the pairwise interactions between particles in the base cell and their neighbor cells \cite{Gratl_Autopas_2022}}
    \label{fig:basesteps}
\end{figure}
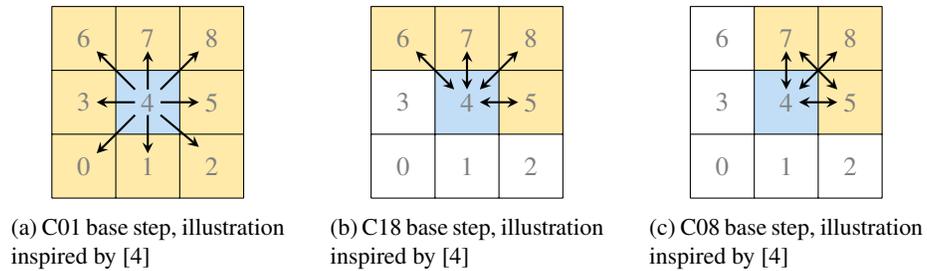

Here, AutoPas distinguishes between three different base steps for the traversals, highlighted in \autoref{fig:basesteps}.
The base step defines the operation carried out on a single cell, i.e., with which neighboring cell the interactions take place.
Furthermore, it poses constraints on which cells the base step can be carried out simultaneously, as neighboring cells required from two different base steps must not overlap.
This results in a domain coloring scheme where cells of the same color can be processed in parallel. \cite{Gratl_Autopas_2022}

In the end, every traversal passes the particle lists of the interacting cells as input to the force functor that carries out the force calculation.

\begin{figure}[h]
    \centering

    \begin{tikzpicture}
        
        \foreach \x in {0,0.5,1} {
            \foreach \y in {4.25,4.75,5.25} {
                \fill [ownorange!30] (\x,\y) rectangle (\x+0.5,\y+0.5);
                \draw (\x,\y) rectangle (\x+0.5,\y+0.5);
                \draw (\x+0.25,\y+0.25) node {?};
            }
        }
        \fill [ownblue!50] (0.5,4.75) rectangle (1,5.25);
        \draw (0.5,4.75) rectangle (1,5.25);

        \foreach \x in {3.375} {
            \foreach \y in {5.25} {
                \coordinate (A) at (\x,\y);
                \coordinate (B) at (\x+0.1+rand/7,\y+0.3+rand/7);
                \coordinate (C) at (\x+0.3+rand/7,\y+0.2+rand/7);
                \coordinate (D) at (\x+0.4+rand/7,\y-0.1+rand/7);
        
                \draw [dashed, gray!90] (A) -- (B);
                \draw [dashed, gray!90] (B) -- (C);
                \draw [dashed, gray!90] (C) -- (D);
                \draw [dashed, gray!90] (D) -- (A);
        
                \fill [ownblue!90] (A) circle (2pt);
                \draw [gray!85] (A) circle (2pt);
                \fill [ownblue!90] (B) circle (2pt);
                \draw [gray!85] (B) circle (2pt);
                \fill [ownblue!90] (C) circle (2pt);
                \draw [gray!85] (C) circle (2pt);
                \fill [ownblue!90] (D) circle (2pt);
                \draw [gray!85] (D) circle (2pt);
            }
        }

        \foreach \x in {2.85,3.9} {
            \foreach \y in {4.4} {
                \coordinate (A) at (\x,\y);
                \coordinate (B) at (\x+0.1+rand/7,\y+0.3+rand/7);
                \coordinate (C) at (\x+0.3+rand/7,\y+0.2+rand/7);
                \coordinate (D) at (\x+0.4+rand/7,\y-0.1+rand/7);
        
                \draw [dashed, gray!90] (A) -- (B);
                \draw [dashed, gray!90] (B) -- (C);
                \draw [dashed, gray!90] (C) -- (D);
                \draw [dashed, gray!90] (D) -- (A);
        
                \fill [ownorange!90] (A) circle (2pt);
                \draw [gray!85] (A) circle (2pt);
                \fill [ownorange!90] (B) circle (2pt);
                \draw [gray!85] (B) circle (2pt);
                \fill [ownorange!90] (C) circle (2pt);
                \draw [gray!85] (C) circle (2pt);
                \fill [ownorange!90] (D) circle (2pt);
                \draw [gray!85] (D) circle (2pt);
            }
        }

        \draw [-stealth] (2,4.75) -- (2,3.9);
        
        \draw (0,2.75) rectangle (4,3.75);
        \draw (2,3.25) node {Domain Traversal};


        \draw [-stealth] (0.5,2.65) -- (0.5,1.35);
        \fill [white] (0,1.65) rectangle (1,2.35);

        \foreach \x in {0,0.25,0.5,0.75,1,1.25} {
            \draw (\x,1.7) -- (\x,2);
        }

        \draw (2.75-3,1.95) rectangle (4.75-3,2.25);
        \fill [white] (2.625-3,1.825) rectangle (4.625-3,2.125);
        \draw (2.625-3,1.825) rectangle (4.625-3,2.125);
        \fill [white] (-0.5,1.7) rectangle (1.5,2);
        \fill [ownblue!75] (2.5-3,1.7) rectangle (2.75-3,2);
        \draw (2.75-3,1.7) -- (2.75-3,2);
        \fill [ownblue!75] (4.25-3,1.7) rectangle (4.5-3,2);
        \draw (4.25-3,1.7) -- (4.25-3,2);
        \draw (-0.5,1.7) rectangle (1.5,2);
        \scalefont{0.75}
        \scalefont{1.25}

        \foreach \x in {0,0.25,0.5,0.75,1,1.25} {
            \draw (\x,1.7) -- (\x,2);
        }

        \draw [-stealth] (3.5,2.65) -- (3.5,1.35);
        \fill [white] (3,1.65) rectangle (4,2.35);

        \foreach \x in {3,3.25,3.5,3.75,4,4.25} {
            \draw (\x,1.7) -- (\x,2);
        }

        \draw (2.75,1.95) rectangle (4.75,2.25);
        \fill [white] (2.625,1.825) rectangle (4.625,2.125);
        \draw (2.625,1.825) rectangle (4.625,2.125);
        \fill [white] (2.5,1.7) rectangle (4.5,2);
        \fill [ownorange!75] (2.5,1.7) rectangle (2.75,2);
        \draw (2.75,1.7) -- (2.75,2);
        \fill [ownorange!75] (4.25,1.7) rectangle (4.5,2);
        \draw (4.25,1.7) -- (4.25,2);
        \draw (2.5,1.7) rectangle (4.5,2);
        \scalefont{0.75}
        \foreach \x in {3,3.25,3.5,3.75,4,4.25} {
            \draw (\x,1.7) -- (\x,2);
        }
        \scalefont{1.25}

        \draw (0.25,0.25) rectangle (4.25,1.25);
        \fill [white] (0.125,0.125) rectangle (4.125,1.125);
        \draw (0.125,0.125) rectangle (4.125,1.125);
        \fill [white] (0,0) rectangle (4,1);
        \draw (0,0) rectangle (4,1);
        \draw (2,0.5) node {Force Functor};

        \draw [-stealth] (2,-0.15) -- (2,-0.5);
        \draw (2,-0.75) node {Pairwise Forces};
        
    \end{tikzpicture}
    
    \caption{Force calculation control flow of AutoPas. The particle containers carry out the neighbor identification and decide on the data structures for storing the particles. The (parallel) domain traversals pass the concrete particle lists to the force functor that calculates the pairwise forces.}
    \label{fig:flow}
\end{figure}
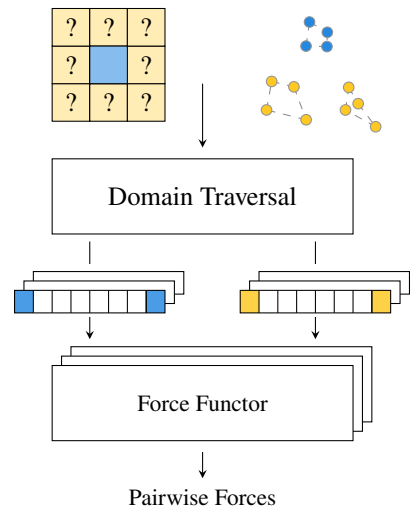

When using the Verlet Cluster List algorithm for neighbor identification, the traversal indicates in which order the clusters are processed.
The functor receives the base cluster as the first list and the neighboring clusters as the second list.
\autoref{fig:flow} highlights AutoPas' force computation control flow.
A more detailed description, especially for the traversal mechanisms, can be found in \cite{Gratl_Autopas_2022}.

\subsubsection{Data Layout}
In AutoPas, the data layout governs how the particle properties are stored in memory.
One option is called Array of Structures (AoS), where particles are represented as C++ objects with properties like positions, forces, etc.
They are then inserted in a C++ \texttt{std::vector<Particle>}, and the whole objects are placed in consecutive memory.
However, this is ill-suited for vectorization as, e.g., the position data would need to be gathered from different memory locations into one register. \cite{Gratl_Early_2019}

The opposite is provided by the Structure of Arrays (SoA) memory layout, where the particles' properties are stored consecutively.
All e.g., x-positions, end up in a C++ vector of \texttt{double} values with one entry for each particle.
This allows for easier transfer from memory into the SIMD registers and vice versa. \cite{Gratl_Early_2019}

\subsubsection{Auto-Tuning strategy}

The preceding sections discussed some supported neighbor identification algorithms, traversals, and the data layout, which form a subset of the possible tunable parameters.
The overall search space for the algorithm selection is then constructed by the cartesian product of all configured parameters, e.g. \texttt{$[\text{Container}] \times [\text{Traversal}] \times [\text{Newton3}] \times ... \times [\text{Data Layout}]$}.

Currently, AutoPas supports a full search approach for selecting the optimal configuration regarding a chosen tuning metric.
Full search assumes that consecutive pairwise force calculation iterations yield comparable execution times.
Therefore, several different configurations are used in successive iterations.
AutoPas users can specify how many samples are drawn for each configuration to achieve more resilient measurements.
The tuning metric's values during the force computation of each configuration are recorded for the number of chosen iterations and reduced to a single value, e.g., average or absolute minimum.
The optimal configuration regarding the chosen evidence is executed for a specified number of iterations before the next tuning phase starts.
This approach is known to be suboptimal, as it has to execute every configuration for a certain time in each tuning phase, including suboptimal configurations.
However, full search can be used to demonstrate the potential of auto-tuning.
\cite{Gratl_Autopas_2022}\cite{Multisite_2023}

\subsubsection{Energy Consumption as Tuning Metric}

Since Gratl et al. \cite{Gratl_Autopas_2022} published the first reference paper, AutoPas now supports measuring the energy consumption for each iteration.
For this purpose, the Power Measurement Toolkit (PMT) \cite{pmt_2022} is used, which allows the sampling of the energy consumption in a specific code region.
The measuring capabilities are independent of the architecture and provide an abstraction from the vendors' hardware-specific APIs.
This allows AutoPas not only to gather data about energy consumption but also to choose it as a tuning metric.

\section{Related Work}
\label{chap:related_work}
After clarifying the relevant background, this chapter provides an overview of other projects that employ vectorization in MD and how they differ from this work.
The popular MD simulators, such as LAMMPS\cite{lammps_tersoff_2016} or GROMACS \cite{Gromacs_simd_2015}, have already proved the potential of SIMD vectorization in their code.
They also investigated the influences of the orders in which the particles are loaded into the registers.

Höhnerbach et al. \cite{lammps_tersoff_2016} suggested considering multiple vectorization schemes for the Tersoff-Multibody-Potential in LAMMPS.
For this type of potential, three nested loops iterate over particles and they investigate different options for which loop to map to sequential execution, parallelism, or vectorization.
They mention that the option choice can depend on the SIMD width and the sizes of the neighbor lists for the Verlet Lists algorithm.
Instead, in this work, other neighbor identification algorithms and more runtime parameters, such as traversals, are investigated.

With GROMACS, another popular simulator considered different schemes for filling the vector registers.
Páll et al. \cite{gromacs_overview_2020} show that GROMACS adjusts the algorithmic settings mainly based on the size of the SIMD width and chooses different patterns for CPU and GPU executions.
Additionally, they support a wider range of options on architectures on which partial vector register access can be realized efficiently.
On CPUs, the same $i$-particle (base cluster) and different $j$-particles (neighboring clusters) are assigned to a pair of SIMD registers for interaction.
On the other hand, their GPU algorithm combines multiple j-clusters to saturate the GPU's large SIMD-like execution units.
In conclusion, the main difference to this paper is their limitation to architecture-specific considerations.

Furthermore, Tchipev \cite{ls1mardyn_2020} also investigated two different vectorization schemes for the Linked Cells algorithm in ls1mardyn.
The schemes differ in which of the nested loops is vectorized.
The considered possibilities are the inner or both loops.
In the end, vectorization over the inner loop is claimed to be better suited for handling multi-site molecules and is statically chosen.

In the scope of energy consumption, Jakobs et al. \cite{power_consumption_2016} explored the impact of vectorization on power and energy consumption for Matrix multiplications.
They find that auto-vectorization can lead to a different power and energy consumption than intrinsics-based SIMD code.
However, their algorithm does not change and the use case of Matrix multiplication is quite generic.
Although the effect of vectorization on energy consumption was already the subject of research, to the best of the authors' knowledge, the different vectorization orders for MD simulations have not yet been examined regarding this metric.

In conclusion, other popular simulators choose their vectorization schemes to be optimal for a few specific use cases, mainly depending on the architecture or algorithm.
Consequently, this work differs as it extends AutoPas, with the possibility of tuning all supported neighbor identification algorithms independently during execution.

\section{SIMD Implementations}
\label{chap:implementation}

The related work indicates that the order in which the particles are loaded into the SIMD registers is important.
AutoPas's extension to the different particle interaction orders is discussed in the following section.
All vector instructions are implemented with the SIMD wrapper library Google Highway\footnote{\url{https://github.com/google/highway}} to abstract from architecture and vector length-dependent intrinsics.

Additionally, the search space for the tuning algorithm is extended to the vectorization patterns.
As this multiplies the search space by the number of vectorization schemes, the full search tuning approach should be used cautiously.
However, it can be used to examine whether the optimal pattern can change over runtime as it tests every possible configuration of parameters at least once.

\subsection{Vectorization Patterns}

In AutoPas, the force functors calculate the intermolecular force.
They receive two lists of particles as input and two nested loops iterate over them.
When applying vectorization, the respective $i$-particle (outer loop) and $j$-particle (inner loop) values must be loaded into the SIMD registers.
Next, the force kernel computes the Lennard-Jones interactions between the pairs of registers.
If requested, Newton's third law needs to be applied in every inner-loop iteration, and at the end of the inner loop, the force values must be stored at the respective particle locations.
A pseudocode algorithm for this procedure can be found in Algorithm \ref{alg:pairwise}.

\begin{algorithm}[h]
    \begin{algorithmic}

        \For {$i \in first\_list$}
            \State i\_forces\_accumulator $\gets 0$
            \State \colorbox{ownorange!35}{i\_reg $\gets$ fill\_i\_registers()}
            \For {$j \in second\_list$}
                \State \colorbox{ownorange!35}{j\_reg $\gets$ fill\_j\_registers()}
                \State forces $\gets$ ForceKernel(i\_reg, j\_reg)
                \State \colorbox{ownorange!35}{applyNewton3(forces, j)} 
                \State i\_forces\_accumulator $\gets$ i\_forces\_accumulator + forces
            \EndFor
            \State \colorbox{ownorange!35}{storeForces(i\_forces\_accumulator, i)} 
        \EndFor
        
    \end{algorithmic}
    \caption{Vectorized Pairwise Force Computation}\label{alg:pairwise}
\end{algorithm}

The structure of both loops already determines the general interaction order of the particles.
So the question arises of how many different $i$- and $j$-particles are loaded into the registers.
The constraint in this paper is that the product of both numbers must equal the vector registers' size to compute as many distinct particle interactions as possible.
When implementing these different interaction orders, the highlighted operations in the pseudo-code in Algorithm \ref{alg:pairwise} must be adapted.
For an 8-way SIMD machine, four patterns arise to compute eight different particle interactions, depending on which loop to vectorize.
This means to take one or two distinct particles from the $i$-loop and eight or four from the $j$-loop, leading to the \texttt{1/8} and \texttt{2/4} patterns.
In both approaches, the loads can be interchanged to the other list resulting in the \texttt{8/1} and \texttt{4/2} patterns.
Their names indicate how many particles are taken from which list, i.e., \texttt{$[$i-particles$]/[$j-particles$]$}.

The expected number of blank entries in the SIMD registers is a criterion to evaluate the patterns.
An attempt to load eight particle values from a list that holds only five will result in three unused entries in the vector registers.
The following paragraphs discuss the filling patterns for the registers and provide a theoretical analysis of the expected number of blank entries.

\subsubsection{Filling Registers}

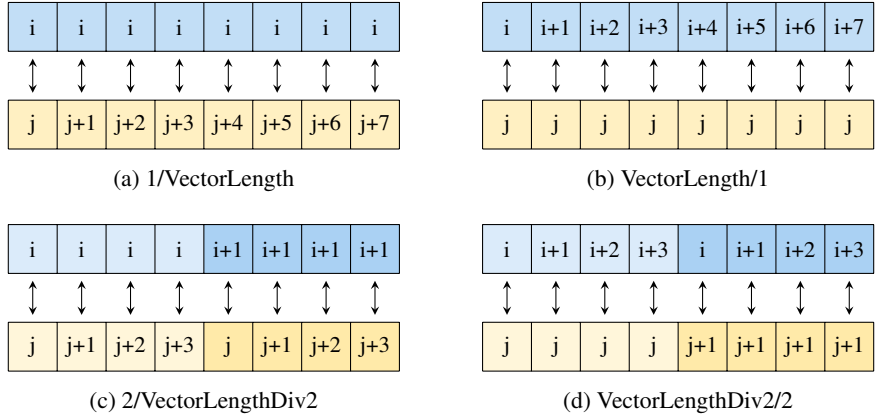
\begin{figure*}
    \centering
    \scalefont{0.85}
    \begin{subfigure}[b]{0.48\textwidth}
        \centering
        \begin{tikzpicture}[scale=0.65]

            \fill [ownblue!25] (0,2) rectangle (8,3);
            \draw (0,2) grid (8,3);
            \foreach \x in {0,1,...,7} {
                \draw (\x+0.5,2.5) node {i};

                \draw [stealth-stealth] (\x+0.5,1.85) -- (\x+0.5,1.15);
            }

            \fill [ownorange!25] (0,0) rectangle (8,1);
            \draw (0,0) grid (8,1);
            \draw (0.5,0.5) node {j};
            \foreach \x in {1,2,...,7} {
                \draw (\x+0.5,0.5) node {j+\x};
            }

        \end{tikzpicture}
        \subcaption{1/VectorLength}
        \label{fig:1xVec}
        
    \end{subfigure}
    \hfill
    \begin{subfigure}[b]{0.48\textwidth}
        \centering
        \begin{tikzpicture}[scale=0.65]

            \fill [ownblue!25] (0,2) rectangle (8,3);
            \draw (0.5,2.5) node {i};
            \foreach \x in {1,2,...,7} {
                \draw (\x+0.5,2.5) node {i+\x};
            }
            \draw (0,2) grid (8,3);

            \fill [ownorange!25] (0,0) rectangle (8,1);
            \foreach \x in {0,1,...,7} {
                \draw (\x+0.5,0.5) node {j};
                \draw [stealth-stealth] (\x+0.5,1.85) -- (\x+0.5,1.15);
            }
            \draw (0,0) grid (8,1);
            
        \end{tikzpicture}
        \subcaption{VectorLength/1}
        \label{fig:Vecx1}
    \end{subfigure}
    \vskip\baselineskip
    \begin{subfigure}[b]{0.48\textwidth}
        \centering
        \begin{tikzpicture}[scale=0.65]

            \fill [ownblue!15] (0,2) rectangle (4,3);
            \fill [ownblue!35] (4,2) rectangle (8,3);
            \draw (0,2) grid (8,3);
            \foreach \x in {0,1,...,3} {
                \draw (\x+0.5,2.5) node {i};
                \draw [stealth-stealth] (\x+0.5,1.85) -- (\x+0.5,1.15);
            }
            \foreach \x in {4,5,...,7} {
                \draw (\x+0.5,2.5) node {i+1};
                \draw [stealth-stealth] (\x+0.5,1.85) -- (\x+0.5,1.15);
            }

            \fill [ownorange!15] (0,0) rectangle (4,1);
            \fill [ownorange!35] (4,0) rectangle (8,1);
            \draw (0,0) grid (8,1);
            \draw (0.5,0.5) node {j};
            \draw (4.5,0.5) node {j};
            \foreach \x in {1,2,3} {
                \draw (\x+0.5,0.5) node {j+\x};
                \draw (\x+4.5,0.5) node {j+\x};
            }

        \end{tikzpicture}
        \subcaption{2/VectorLengthDiv2}
        \label{fig:2xVec2}
    \end{subfigure}
    \hfill
    \begin{subfigure}[b]{0.48\textwidth}
        \centering
        \begin{tikzpicture}[scale=0.65]
            \fill [ownblue!15] (0,2) rectangle (4,3);
            \fill [ownblue!35] (4,2) rectangle (8,3);
            \draw (0.5,2.5) node {i};
            \draw (4.5,2.5) node {i};
            \foreach \x in {1,2,3} {
                \draw (\x+0.5,2.5) node {i+\x};
                \draw (\x+4.5,2.5) node {i+\x};
            }
            \draw (0,2) grid (8,3);

            \fill [ownorange!15] (0,0) rectangle (4,1);
            \fill [ownorange!35] (4,0) rectangle (8,1);
            \foreach \x in {0,1,...,3} {
                \draw (\x+0.5,0.5) node {j};
                \draw [stealth-stealth] (\x+0.5,1.85) -- (\x+0.5,1.15);
            }
            \foreach \x in {4,5,6,7} {
                \draw (\x+0.5,0.5) node {j+1};
                \draw [stealth-stealth] (\x+0.5,1.85) -- (\x+0.5,1.15);
            }
            \draw (0,0) grid (8,1);
            
        \end{tikzpicture}
        \subcaption{VectorLengthDiv2/2}
        \label{fig:Vec2x2}
    \end{subfigure}
    
    \caption{Vector register allocation for the different vectorization patterns for an 8-way SIMD machine. The upper blue registers represent the allocation of the i-particles obtained by taking particles from the outer loop. The lower orange registers visualize the allocation of the j-particles taken from the inner loop.}
    \label{fig:vec_patterns}
\end{figure*}

The \texttt{1/VectorLength} pattern broadcasts one $i$-particle and loads the full vector length from the second list with $j$-particles.
This is AutoPas', ls1mardyn's, and GROMACS' current choice for vectorization when running on CPUs.
The resulting registers are highlighted in \autoref{fig:1xVec}.
With this procedure, one iteration of both nested loops computes the pairwise forces $F_{ij}, F_{i,j+1}, ..., F_{i,j+v-1}$ with $v$ as the vector length.
The drawbacks of this approach are evident for small second lists where the register space can not be utilized completely.
Consequently, the vectorization efficiency will be low because of many blank entries in the SIMD registers.

A reversed approach is provided by the \texttt{VectorLength/1} pattern where the whole SIMD width is loaded with $i$-particles and interacts with one $j$-particle.
The resulting register allocation is visualized in \autoref{fig:Vecx1}.
Therefore, the forces $F_{ij}, F_{i+1,j}, ..., F_{i+v-1,j}$ are calculated in one double-for-loop iteration.
This idea will likely perform well on big first and small second lists.

The \texttt{2/VectorLengthDiv2} (\autoref{fig:2xVec2}) and \texttt{VectorLengthDiv2/2} (\autoref{fig:Vec2x2}) patterns aim at tackling the potential issue of low vectorization efficiency.
They are adapted from the considerations by Höhnerbach et al. \cite{lammps_tersoff_2016} and Páll et al. \cite{gromacs_overview_2020} and require efficient loading/storing of sub-registers.
With this procedure, fewer blank entries in the registers are expected, especially when the interaction buffers are small.
However, populating the registers gets more complicated as simple Load- and Broadcast-instructions no longer suffice.

\subsubsection{Store Force Contribution}

Once the inner loop computes all force contributions for one (or more) $i$-particles, their values must be added to the correct memory locations.

For the \texttt{1/VectorLength} order, only one $i$-particle is considered.
Consequently, the task is to sum up all values in the force register and add them to the force at the memory location of the particle with index $i$.
With \texttt{ReduceSum()}, Highway provides an efficient instruction to sum all register values.
The resulting store operation does not need to be vectorized, as only one location is accessed.

The \texttt{2/VectorLengthDiv2} approach follows a similar idea.
However, the force contributions in the lower half of the register belong to the $i$ particle and the upper half to the $i+1$ particle.
As a result, the register is logically divided, and the lower values are summed up independently from the upper values.
Highway provides abstract functions for this with \texttt{LowerHalf()} and \texttt{UpperHalf()}.
However, the underlying architecture is not guaranteed to implement these generic functions efficiently.
The add and store operations are then carried out sequentially at the corresponding memory location of the particles $i$ and $i+1$.

In the case of the \texttt{VectorLength/1} pattern, the resulting force register holds distinct $i$-particle force values.
Therefore, no reduction has to be applied, and only the store operation is vectorized to add the register values to the memory locations $i$, $i+1$, ..., $i+v-1$.

Similarly, the \texttt{VectorLengthDiv2/2} idea holds multiple $i$-contributions.
However, the particle contributions in the lower half of the registers match the contributions of the upper half.
Consequently, both halves of the register must be accumulated, and the resulting values can be added to the value at the memory locations $i$, $i+1$, ..., $i+v/2-1$.
The current implementation relies on the highway functions \texttt{LoadN()} and \texttt{StoreN()} instead of masked operations to load and store half of the full register size.
Whether this can be executed efficiently can depend on the underlying architecture.

\subsubsection{Newton's Third Law}

Newton's third law is already mentioned earlier as a possibility to halve the number of force calculations.
The pairwise force $F_{ij}$ is accumulated to the $i$ particle's force and subtracted from the $j$ particle's force.
Consequently, another memory access must be considered and aligned with the particle interaction orders for each force calculation.

Compared to the usual force accumulation for the $i$-particles, Newton's third law is applied to every calculated force contribution for $j$-particles in the inner loop.
Consequently, for the \texttt{1/VectorLength} pattern, a vectorized subtraction on the $j$-particles' force memory locations has to be carried out.
Here, the same instructions can be utilized compared to the \texttt{VectorLength/1} order for storing the force contributions to the $i$-particles.
In general, access to the $j$-particles when handling Newton's third law follows the same operations when accumulating the $i$-particle forces.

\section{Performance Benchmarks}
\label{chap:results}

After discussing the implementation of the considered improvements, the next chapter discusses the benchmark settings and their results.

AutoPas is not a complete simulator but focuses on providing high performance for computing short-range particle interactions.
Therefore, it ships with its own simulation code called md-flexible\footnote{\url{https://github.com/AutoPas/AutoPas/tree/master/examples/md-flexible}}.
It is a simple program that allows for easy testing and benchmarking of AutoPas' algorithms. \cite{Gratl_Autopas_2022}

All discussed benchmarks are carried out on the WindHPC cluster\footnote{\url{https://www.windhpc.de/}} of the Helmut Schmidt University in Hamburg.
It provides Intel Xeon Platinum 8260L processors with AVX512 SIMD support.
Furthermore, it offers CMake 3.26.5 and GCC 11.4.1-3.
Each compute node features two sockets with 24 cores per socket and two threads per core, resulting in 96 available hardware threads per compute node.

\subsection{Benchmark A: Exploding Cube}

The first toy benchmark only utilizes 16 OpenMP Threads as the chosen setting is with 5,920 particles comparably small.
Here, a cube of densely packed particles explodes and spreads over the simulation domain (200x200x200) for 60,000 iterations with reflective boundaries on each side.
A complete definition of the configuration can be found in the AutoPas repository\footnote{examples/md-flexible/input/vectorizationBenchmarking/patternTests/Exploding\_Cube\_*.yaml (varying suffix for Linked Cells and Verlet Cluster Lists), Commit ID: a925340}
Consequently, the density of the particles is high at the beginning and decreases over time, which is expected to affect the number of particles per cell and the sizes of the neighbor lists.

\begin{figure}

    \begin{subfigure}[b]{0.49\textwidth}
        \includegraphics[width=\textwidth]{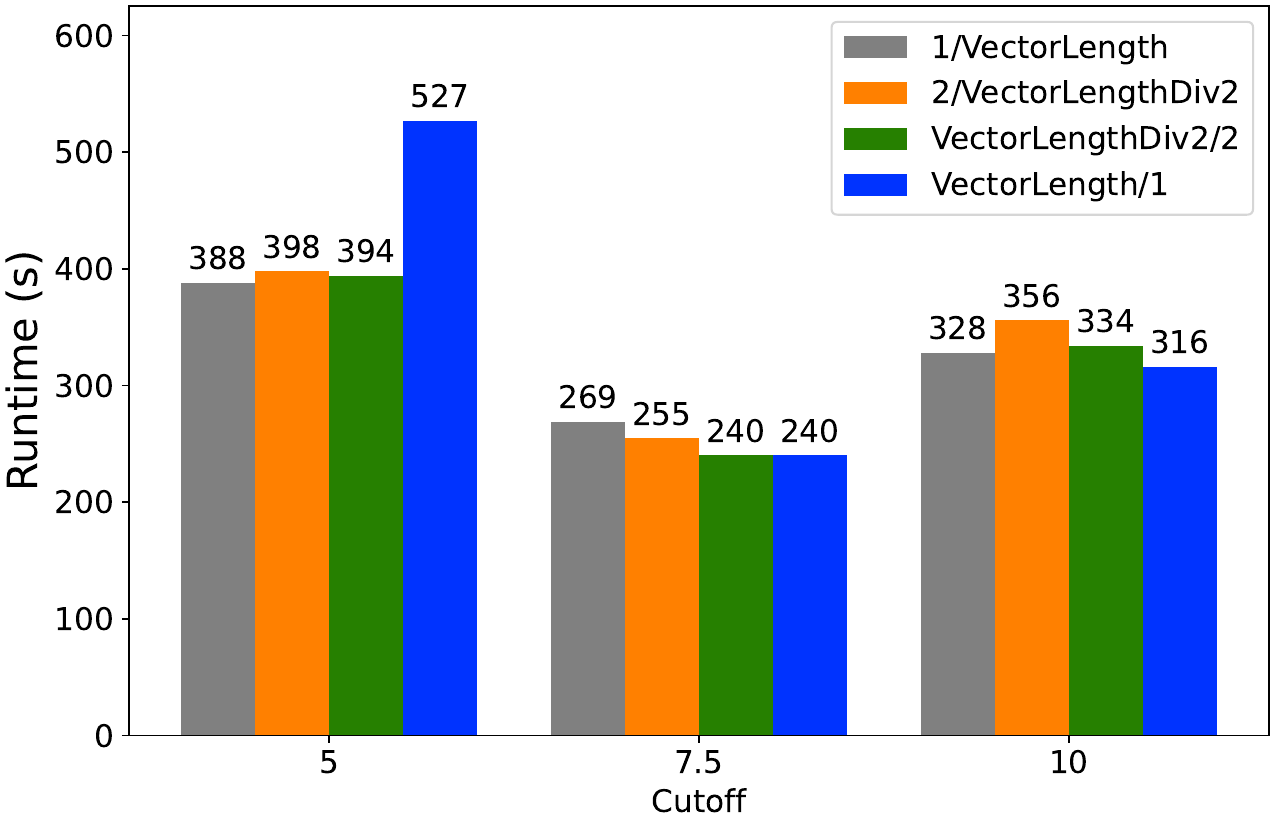}
        \caption{Different cutoff distances, Domain Traversal: \texttt{lc\_c18}, Newton3 optimization: \texttt{disabled}, container: Linked Cells}
        \label{fig:cutoff_change}
    \end{subfigure}
    \hfill
    \begin{subfigure}[b]{0.49\textwidth}
        \includegraphics[width=\textwidth]{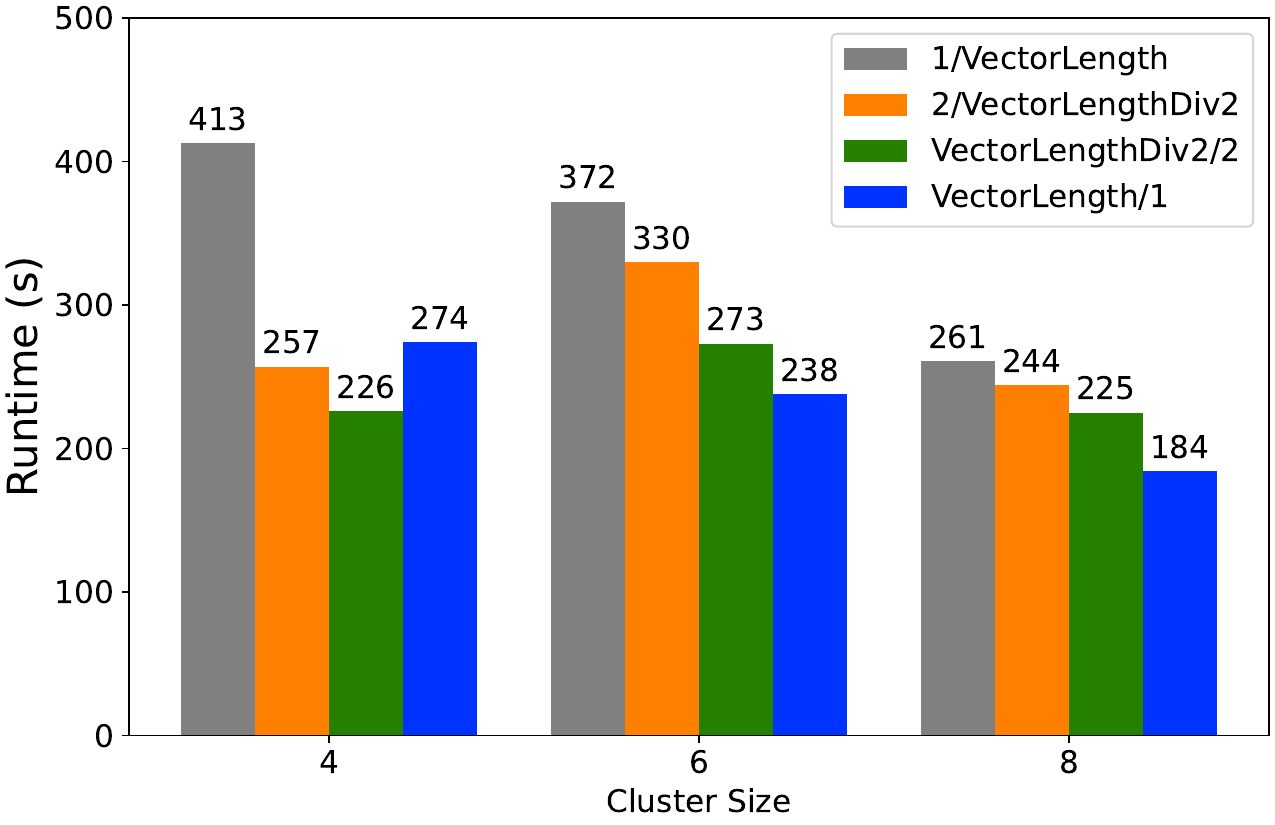}
        \caption{Different cluster sizes, Domain Traversal: \texttt{vcl\_c06}, Newton3 optimization: \texttt{enabled}, Container: Cluster Lists, cutoff: 5}
    \label{fig:cluster_change}
    \end{subfigure}

    \vskip\baselineskip

    \begin{subfigure}[b]{0.49\textwidth}
        \includegraphics[width=\textwidth]{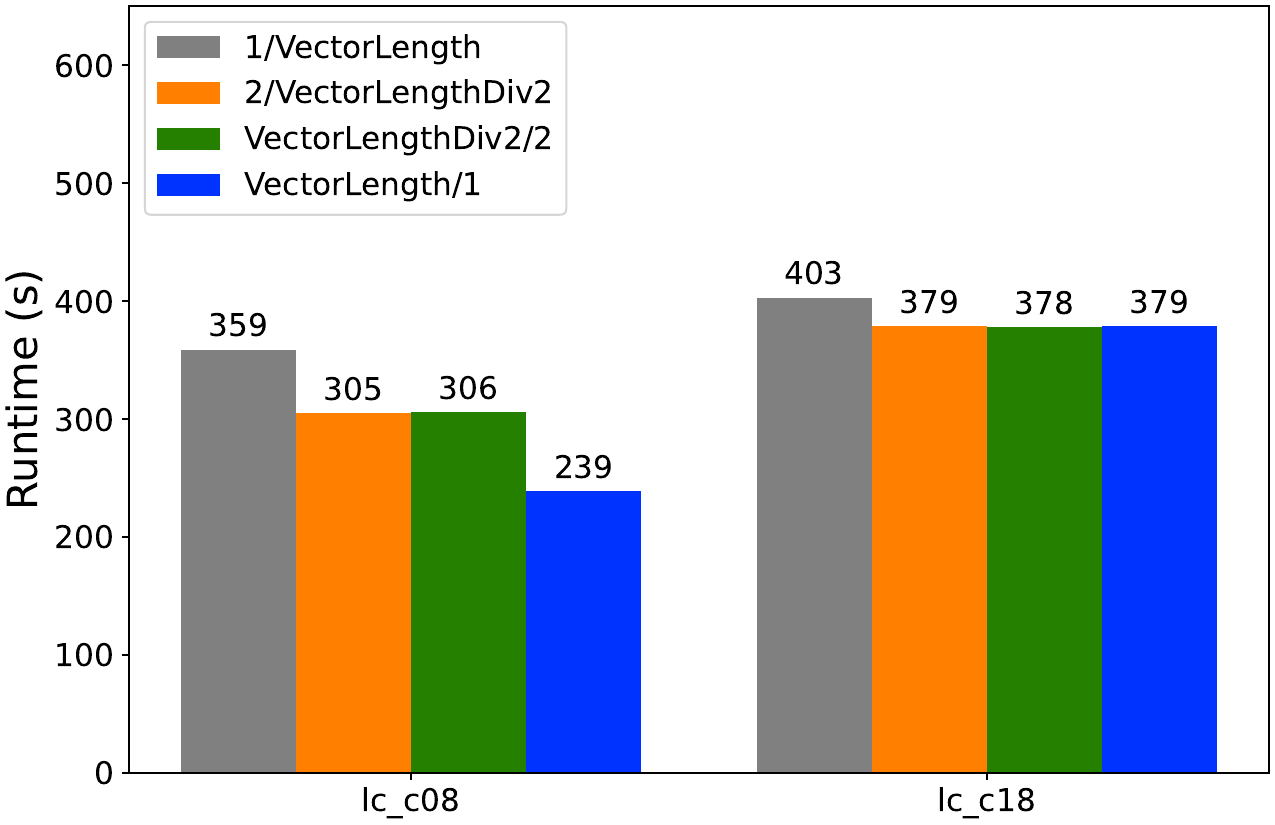}
        \caption{Differernt domain traversals, Newton3: \texttt{enabled}, container: Linked Cells, cutoff: 5}
        \label{fig:traversal_change}
    \end{subfigure}
    \hfill
    \begin{subfigure}[b]{0.49\textwidth}
        \includegraphics[width=\textwidth]{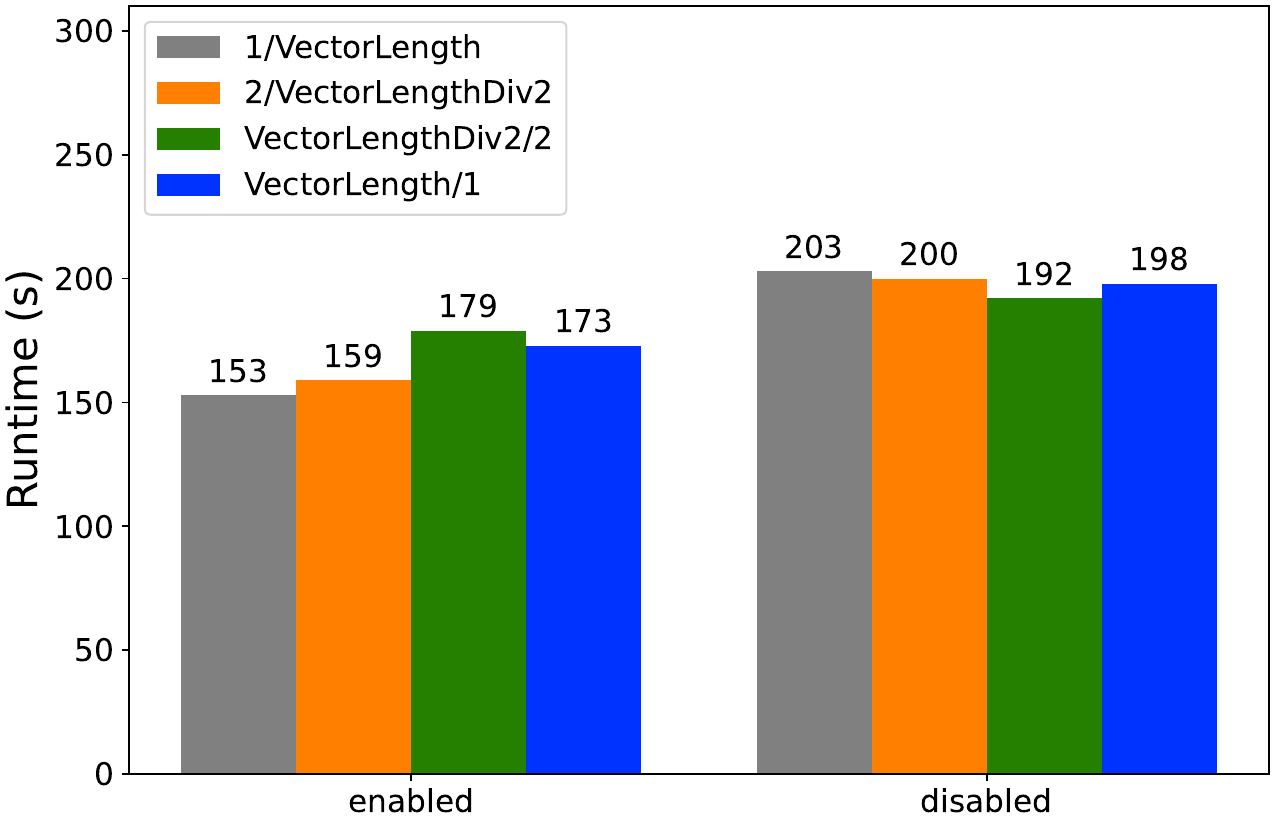}
        \caption{Different Newton3 settings, Domain Traversal: \texttt{lc\_c08}, container: Linked Cells, cutoff: 7.5}
        \label{fig:newton_change}
    \end{subfigure}

    \caption{Multiple different comparisons of parameters that affect the choice of the optimal vectorization pattern.}
    \label{fig:toy_differences}
\end{figure}

Each vectorization pattern is independently tested for different parallel domain traversals, neighbor identification algorithms, algorithm-specific settings such as the cluster size, and simulation-specific settings as the cutoff.
The goal is to derive a rough understanding of how to guide auto-tuning in the future.
The runtimes are exclusively measured for the force calculation, including traversing the simulation domain and invocating the force kernel.
With this procedure, the parts affected by the vectorization patterns can be examined in isolation.
\autoref{fig:toy_differences} provides runtime plots depending on multiple different parameters such as the cutoff distances, cluster sizes, parallel domain traversals, and settings of the Newton3 optimization.

\subsubsection{Cutoff Difference}
The graph in \autoref{fig:cutoff_change} highlights that the choice of the best-performing vectorization order can change depending on the configured cutoff distance for the Linked Cells container.
On the one hand, the \texttt{VectorLength/1} takes up to 25\% more execution time than the other orders for smaller cutoff sizes but outperformed them for bigger cutoff distances by up to 12\%.
As the cutoff distance influences the cell sizes and the number of particles per cell, this result indicates that the sizes of the interaction buffers can affect the optimal performance of the vectorization schemes.

It has to be noted that in this experiment, a decreasing cutoff distance can lead to a higher runtime.
This can be counter-intuitive at first glance, as a smaller cutoff means the relevant interaction distance is smaller.
However, for the Linked Cells algorithm, a smaller interaction distance results in smaller cells and, consequently, more cells over the domain.
At the beginning of the simulation, only a few cells hold many particles, which can lead to a high number of empty cells that need to be traversed.
Hence, this can lead to a considerable load balance between threads with empty cells and threads with populated cells.

On the contrary, the Cluster Lists algorithm does not yield higher runtimes for lower cutoff distances, which supports the assumption of the cells' load imbalance.
Here, the empty parts of the domain are not traversed in much detail, as no particles reside in the cluster lists.
However, the Cluster Lists algorithm is not sensitive to different cutoff sizes regarding the optimality of the various SIMD interaction orders.

\subsubsection{Cluster Size Difference}
This is because the sizes of neighbor lists are always multiples of the cluster size by design.
Whereas the number of clusters per neighbor list can change (based on the particle density and cutoff distance), their alignment remains constant.
\autoref{fig:cluster_change} visualizes the runtime for the various vectorization schemes and cluster sizes of 4, 6, and 8.
The results show that aligning the interaction order to the base cluster size yields the highest performance.
On the 8-way SIMD benchmark machine and a cluster size of 4, the \texttt{VectorLengthDiv2/2} pattern loads the full base cluster only once and interacts it with all neighboring clusters.
Hence, it can achieve a runtime advantage of up to 45\% to the other patterns.
For a cluster size of 8, the cluster size is equal to the SIMD width.
Consequently, it is guaranteed that loading from the base cluster does not lead to blank entries in the vector registers.
Following this reasoning, the performance advantage of up to 30\% compared to the other patterns is not surprising.
Consequently, the size of the base cluster is crucial for the choice of the best-performing pattern in the Cluster Lists algorithm.

\subsubsection{Domain Traversal Difference}
Until now, only parameters that can influence the sizes of the interaction buffers have been discussed.
However, the results in \autoref{fig:traversal_change} for the Linked Cells algorithm show that the optimality of the schemes can depend on the order in which the traversals process the domain.
Most SIMD schemes have a similar runtime for the \texttt{lc\_c18} domain traversal.
Compared to the other orders, a higher runtime of up to 5\% is observed for the \texttt{1/VectorLength} pattern.
In the case of the \texttt{lc\_c08} traversal, the \texttt{VectorLength/1} scheme can achieve the lowest execution time with a much higher runtime difference between 22\% and 33\% compared to the other patterns.

As all different traversal mechanisms compute the same particle interactions, their main difference lies in the order in which the cells' lists interact in the base steps.
This can influence the order in which the cells' particle lists are passed to the force functor.
A list that might have been the $i$-particle list for the \texttt{lc\_c08} can end up being the $j$-list in the \texttt{lc\_c18} case.
Consequently, the sizes of the interaction buffers might have been swapped, leading to a different vectorization pattern being optimal.

Additionally, the base step of both traversals differs, as depicted earlier in \autoref{fig:basesteps}.
For the \texttt{lc\_c18} base step, other cells interact compared to the \texttt{lc\_c08} step, resulting in potentially different memory access patterns and different computational intensities.
These can influence the behavior of the vectorization patterns, as the $i$- and $j$-lists might be accessed in a different fashion.

In the provided figure, the experimental settings favor the \texttt{VectorLength/1} pattern for the \texttt{lc\_c08} traversal.
In other settings (cutoff 7.5 and disabled \texttt{Newton3}, the \texttt{VectorLengthDiv2/2} scheme was the optimal choice for the \texttt{lc\_c08} traversal.
In conclusion, the results for the traversals show that the optimal choice of the SIMD interaction orders can depend on the domain traversals and there is no clear winner for every setting.

\subsubsection{Newton3 Difference}
In addition to the processing order of the cells, enabling or disabling the \texttt{Newton3} optimization can also affect the choice of the optimal vectorization scheme.
In \autoref{fig:newton_change}, the runtimes for the different patterns depending on the \texttt{Newton3} optimization are visualized.
The plot shows that the \texttt{1/VectorLength} pattern is the most optimal, with an advantage of up to 15\% over the other patterns with \texttt{Newton3} enabled.
When turning off the use of \texttt{Newton3}, the difference among the runtimes of the vectorization schemes decreases.
Additionally, the \texttt{VectorLengthDiv2/2} can change from the worst choice to the best choice with a slight performance improvement of up to 5\% compared to the other schemes.

Enabling \texttt{Newton3} means that the pairwise interactions between cells are only carried out in one direction.
As discussed earlier, the force $F_{ij}$ value is also stored for each $j$-particle.
Disabling this optimization leads to the cells interacting in both ways, meaning that a cell's particle list will be passed once as $i$-particles and once as $j$-particles to the force functor.
This can destroy considerations on the sizes of the lists as the benefits of one big and one small particle buffer can vanish when computing the interactions in both directions.

Again, the same observations on the change of optimality are not recorded for the Cluster Lists.
Here, \texttt{Newton3} does not interchange the order in which the clusters interact.
Enabling the optimization states that the cluster $k$ does not end up in the neighbor list of cluster $z$, if cluster $z$ is already in the list of cluster $k$.
Hence, the sizes of the neighbor lists can be influenced.
The previous results show that the choice of the vectorization pattern usually depends on the size of the base cluster.
Consequently, smaller or bigger neighbor lists governed by the \texttt{Newton3} optimization do not affect the optimal choice of the patterns.

In conclusion, the optimality of the patterns for the Cluster Lists algorithm mainly depends on the cluster sizes.
Other runtime-dependent parameters do not contribute to changes in the choice of the optimal vectorization scheme.
In the Linked Cells case, multiple factors that potentially change during runtime affect the choice of the best-performing particle interaction order.
Consequently, the results from this experiment suggest further investigation of a dynamic auto-tuning approach for the Linked Cells container.

\subsection{Benchmark B: Exploding Liquid}

The second scenario follows a similar structure as before.
A dense group of particles of size 45x25x45 explodes and the particles spread over the tube-like simulation domain of size 50x120x50 with periodic boundaries on each side for 50,000 iterations.
The exact configuration can be found in the AutoPas repository\footnote{examples/md-flexible/input/vectorizationBenchmarking/patternTests/explodingLiquid\_hwy.yaml, Commit ID: d15ffc0}.
The domain is decomposed into four equally sized subdomains and assigned to separate MPI ranks, each running with 24 OpenMP threads.
This leads to four independent AutoPas instances, each with its own tuning phases and potentially different optimal configurations.
The search space contains the Linked Cells container, \texttt{lc\_c01} and \texttt{lc\_c08} domain traversals, and \texttt{Newton3} enabled and disabled, resulting in three different configurations per tuning phase (\texttt{lc\_c01 can only work without \texttt{Newton3}}.
These settings are chosen as the final search space as previous executions favored them frequently.
Ten tuning samples are collected for each configuration in each tuning phase, and the value is reduced to the fastest mean value.
Two separate experiments are carried out using energy consumption and the fastest-time-to-solution tuning metrics.

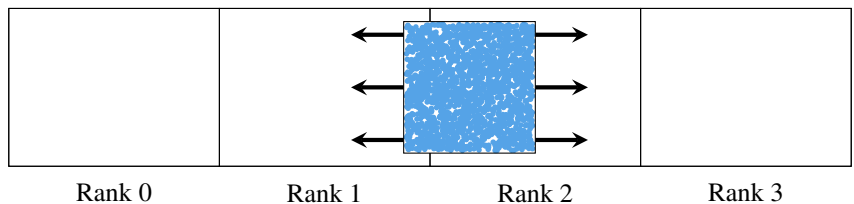
\begin{figure}
    \centering

    \begin{tikzpicture}[scale=0.7]

        \foreach \x in {0,1,2,3} {
            \draw (\x*4,0) rectangle (\x*4+4,3);
            \draw (\x*4+2,-0.5) node {Rank \x};
        }

        \fill[white] (7.5,0.25) rectangle (10,2.75);
        \pgfmathsetseed{1}
        \foreach \x in {1,2,...,2000} {
            \fill[ownblue!70] (7.5+1.25+1.2*rand,1.5+1.2*rand) circle (1.5pt);
        }
        \draw (7.5,0.25) rectangle (10,2.75);

        \foreach \y in {0.5,1.5,2.5} {
            \draw [-stealth, ultra thick] (7.5,\y) -- (6.5,\y);
            \draw [-stealth, ultra thick] (10,\y) -- (11,\y);
        }

        \draw (0,0) rectangle (16,3);
        
    \end{tikzpicture}
    
    \caption{2D sketch of the Exploding Liquid experiment at iteration 0.}
    \label{fig:liquid_init}
\end{figure}

\begin{figure}
    \centering

    \begin{tikzpicture}[scale=0.7]

        \foreach \x in {0,1,2,3} {
            \draw (\x*4,0) rectangle (\x*4+4,3);
            \draw (\x*4+2,-0.5) node {Rank \x};
        }

        \foreach \x in {1,2,...,1000} {
            \fill[ownblue!70] (0.8+0.775*rand,1.525+1.45*rand) circle (1.5pt);
            \fill[ownblue!70] (14.75+1.225*rand,1.525+1.45*rand) circle (1.5pt);
        }
        \foreach \x in {1,2,...,400} {
            \fill[ownblue!70] (8+7*rand,1.525+1.45*rand) circle (1.5pt);
        }
        \foreach \x in {1,2,...,300} {
            \fill[ownblue!70] (7.5+7*rand,2.825+0.125*rand) circle (1.5pt);
            \fill[ownblue!70] (7.5+7*rand,0.175+0.125*rand) circle (1.5pt);
        }
        \foreach \x in {1,2,...,40} {
            \fill[ownblue!70] (2+0.5*rand,2.5+0.25*rand) circle (1.5pt);
            \fill[ownblue!70] (13+0.5*rand,0.5+0.25*rand) circle (1.5pt);
            \fill[ownblue!70] (2+0.75*rand,0.4+0.2*rand) circle (1.5pt);
        }


        \draw (0,0) rectangle (16,3);
        
    \end{tikzpicture}
    
    \caption{2D sketch of the Exploding Liquid experiment at iteration 25,000.}
    \label{fig:liquid_middle}
\end{figure}
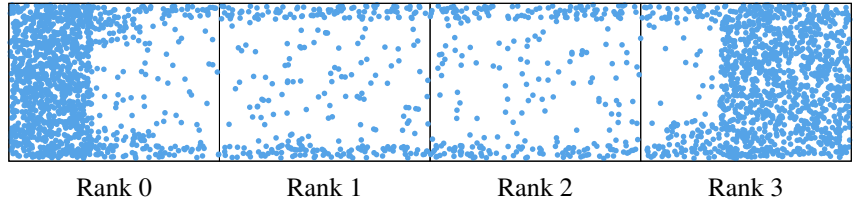

Rank 1 and Rank 2 contain the particle group at the initialization.
The cube is not placed exactly in the middle of the simulation domain to provoke a different behavior on Rank 1 and Rank 2 due to the non-symmetry.
The particles traverse the domain and group again at the left and right boundary at Rank 0 and Rank 3.
A 2D sketch (not scaled to the exact settings) of the simulation behavior is provided in \autoref{fig:liquid_init} and \autoref{fig:liquid_middle}.

On the one hand, the changing particle distribution over time is expected to affect the sizes of the neighbor buffers and, hence, the optimal choice of the SIMD pattern.
Additionally, as the particles spread over the domain in an inhomogeneous fashion, the different subdomains are assumed to lead to different optimal interaction order choices.

\begin{figure}[h]

    \begin{subfigure}[b]{0.49\textwidth}
        \includegraphics[width=\textwidth]{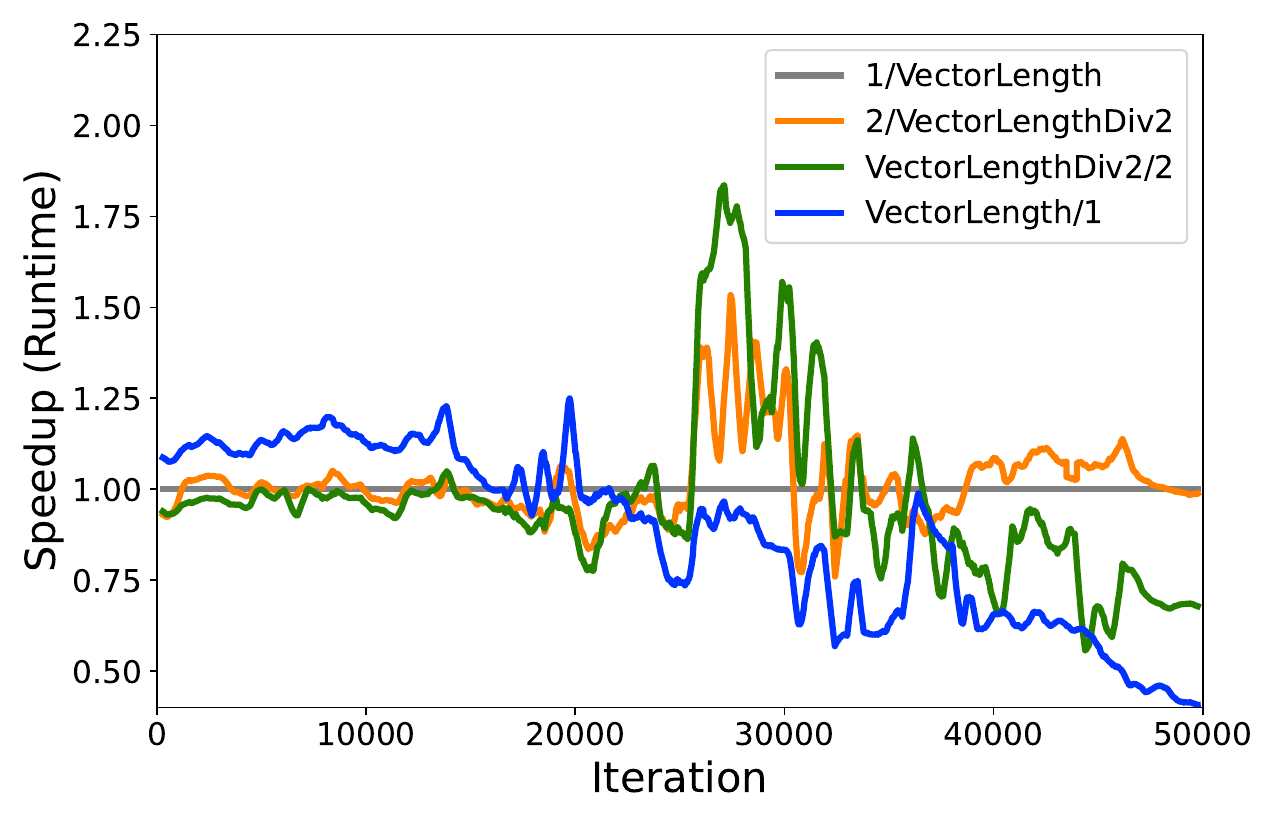}
        \caption{Rank 1, Runtime Speedup}
        \label{fig:r1_runtime}
    \end{subfigure}
    \hfill
    \begin{subfigure}[b]{0.49\textwidth}
        \includegraphics[width=\textwidth]{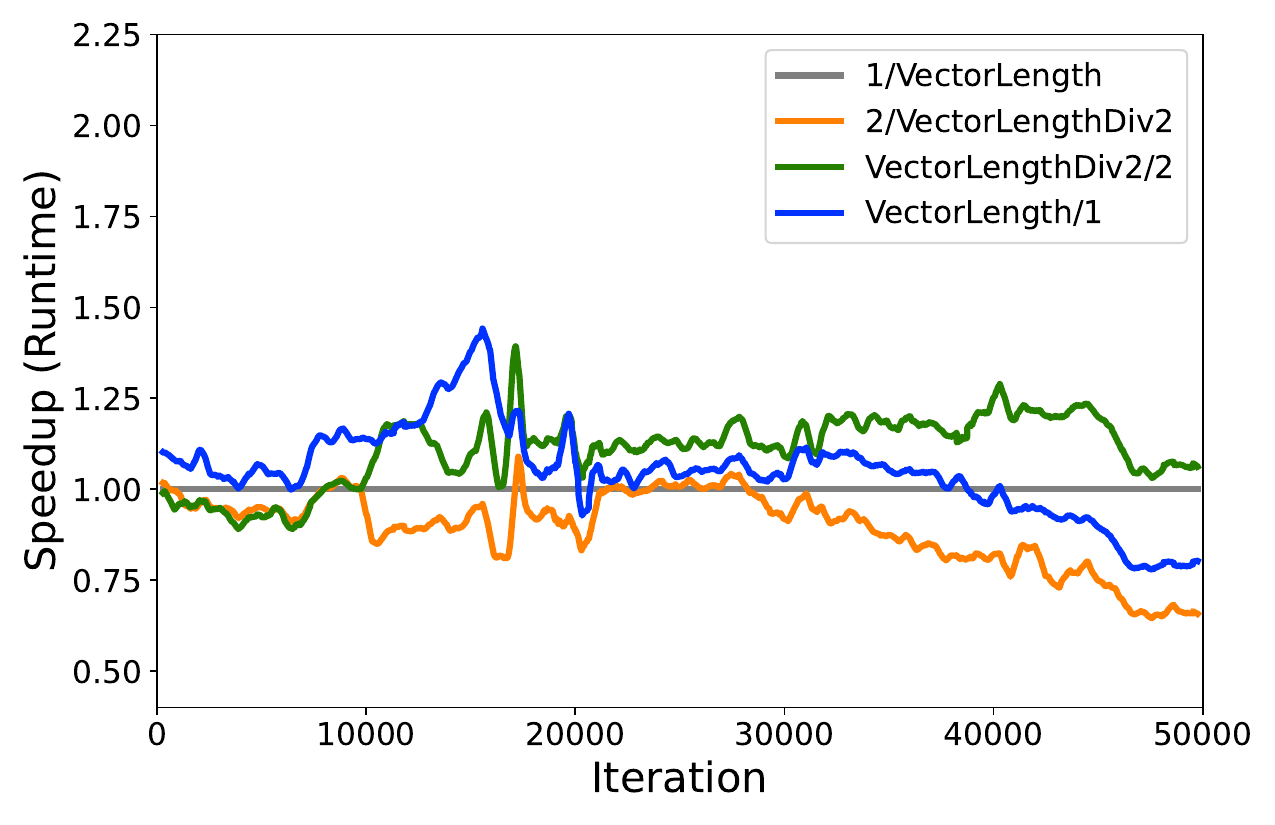}
        \caption{Rank 2, Runtime Speedup}
        \label{fig:r2_runtime}
    \end{subfigure}

    \vskip\baselineskip

    \begin{subfigure}[b]{0.49\textwidth}
        \includegraphics[width=\textwidth]{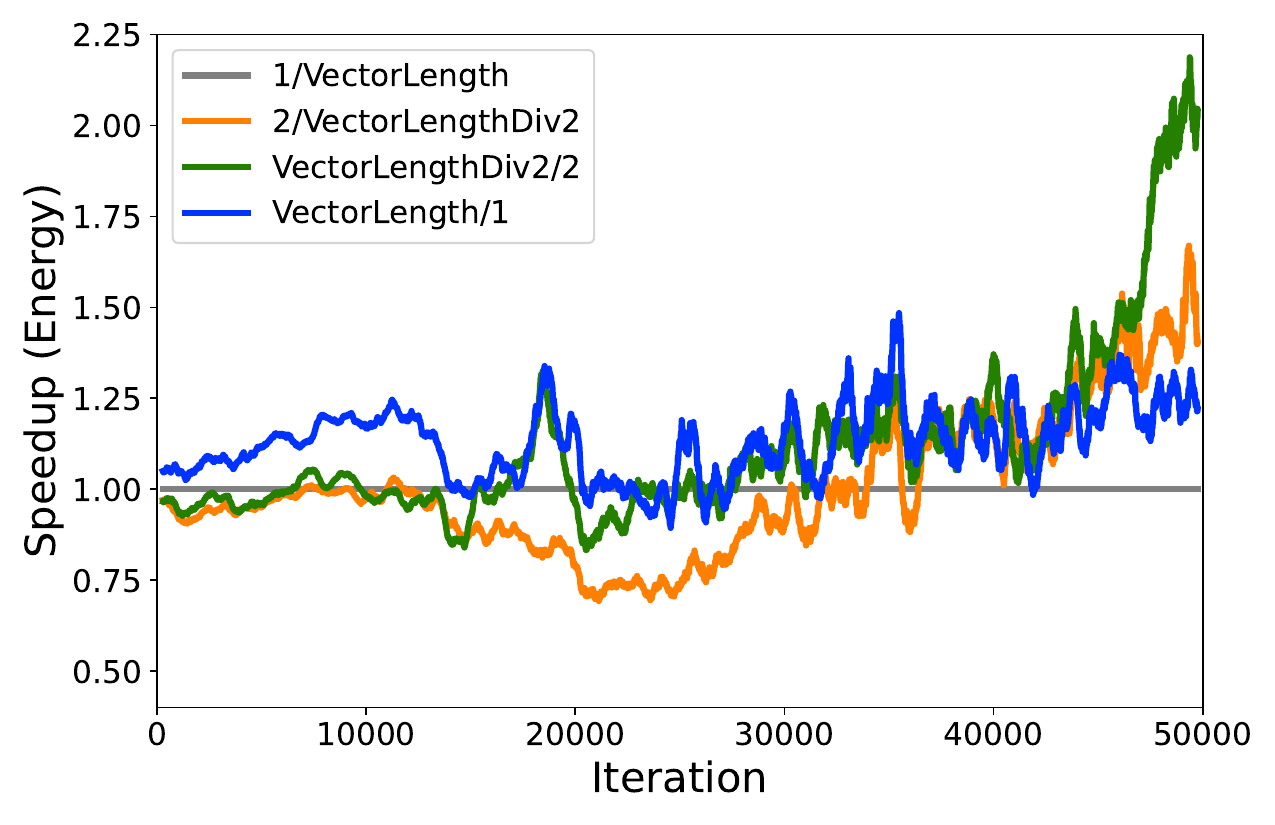}
        \caption{Rank 1, Energy Speedup}
        \label{fig:r1_energy}
    \end{subfigure}
    \hfill
    \begin{subfigure}[b]{0.49\textwidth}
        \includegraphics[width=\textwidth]{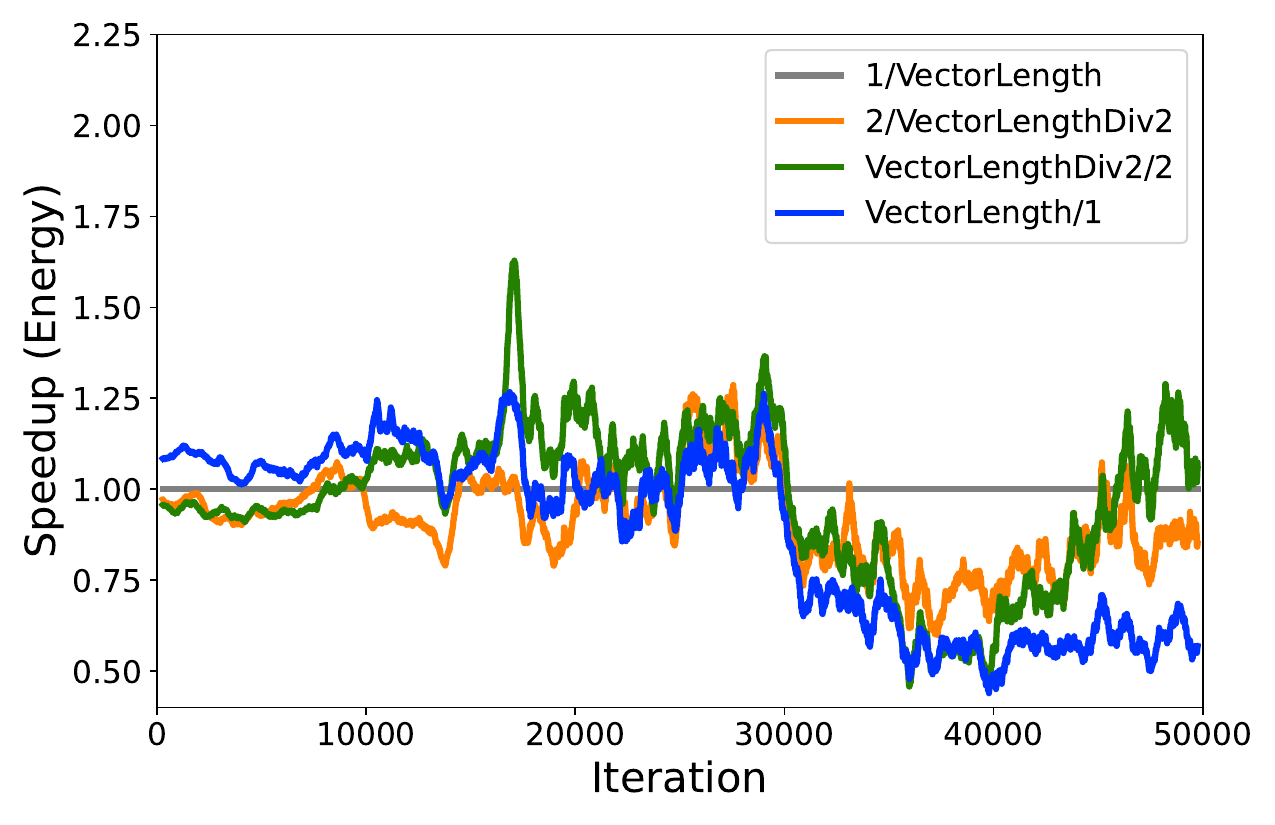}
        \caption{Rank 2, Energy Speedup}
        \label{fig:r2_energy}
    \end{subfigure}

    \caption{Runtime and Energy Speedup plots for both innermost MPI ranks. The data is smoothed with pandas.Dataframe.rolling(window=500).mean().}
    \label{fig:liquid_results}
\end{figure}

This time, the speedup of the interaction orders compared to the \texttt{1/VectorLength} pattern is calculated by \texttt{$\text{Speedup(pattern)} = \text{Metric(1/VectorLength)} \div \text{Metric(pattern)}$}.
Consequently, a high speedup value indicates good performance compared to the reference pattern.
This allows for a better comparison of the patterns' performance w.r.t a chosen metric (execution time/energy consumption) over the runtime of the simulation.
The results visualized in \autoref{fig:liquid_results} show the energy and runtime speedup for the two innermost MPI ranks.
For the outermost patterns, no real change in the optimality of the patterns could be observed as the density of the particles remains constant after the particles group at the edges of the simulation domain after iteration 15,000.

The first observation, which generalizes to both ranks and tuning metrics, is that the optimal vectorization scheme can change during runtime.
Hence, the changing particle distribution over time affects the optimal configuration.
As both ranks lose particles over time, the cells hold fewer particles, leading to a shift in the optimal pattern toward the splitting patterns.

Furthermore, a comparison of the runtime speedup of Rank 1  in \autoref{fig:r1_runtime} and Rank 2 in \autoref{fig:r2_runtime} shows that the different ranks yield different optimal patterns at the end of the simulation.
For Rank 1, the \texttt{2/VectorLengthDiv2} scheme has slightly the highest speedup of around 1.05, whereas, on Rank 2, the \texttt{VectorLengthDiv2/2} order achieves the best runtime with a speedup of up to 1.25.
Even though the general simulation behavior seems identical, both subdomains contain a slightly different number of particles, and the distribution is also different.
This can lead to different memory access patterns being optimal and results in different optimal SIMD interaction orders.

Another interesting behavior of the patterns yields the difference in the optimal runtime and energy schemes.
When choosing runtime as the tuning metric (\autoref{fig:r1_runtime}), the \texttt{2/VectorLengthDiv2} is selected as a slightly optimal pattern.
However, switching the tuning goal to energy consumption changes the optimal pattern to the \texttt{VectorLengthDiv2/2} with a speedup of up to 2.2.
It is observed that the tuning algorithm chooses the same traversal mechanisms for energy and runtime, which indicates that the difference results purely from the vectorization orders.

The exact reasons for this difference can only be speculated as the WindHPC cluster does not provide hardware metrics like memory and cache behavior.
However, the patterns use different higher-level vector instructions that are potentially mapped to different hardware instructions.
It is assumed that the runtime cost of these instructions is not proportional to their energy consumption.
Additionally, the different particle access patterns can save potentially more energy than time and vice versa due to various memory effects.
An in-depth analysis of the exact energy consumption of the used instructions can be subject to further research.

Ultimately, the exploding liquid experiment showed that the different subdomains can yield different optimal vectorization patterns due to a changing particle distribution, which can also change at runtime.
Furthermore, the best-performing algorithm differs when tuning for runtime or energy consumption.
Consequently, the scenario highlights the necessity of AutoPas' auto-tuning for the vectorization orders.

\section{Future Work}
\label{chap:future}

Although the results indicate the necessity of dynamic auto-tuning, the full-search tuning approach is known to be suboptimal because it has to execute every configuration at least once in every tuning phase.
Consequently, future work could incorporate the observations of the circumstances under which the individual patterns are optimal in some knowledge-based or data-driven tuning approach.

Additionally, in this work, only one SIMD architecture was considered.
In the future, other instruction sets, such as ARM's SVE, can be the subject of further investigations.
The potentially different costs for some vector instructions might change the optimal behavior of the interaction orders.

Furthermore, the related work showed that the best interaction patterns can differ when running the force calculation on GPUs.
If AutoPas also considers GPU support in the future, it can be worth investigating various particle interaction orders there as well.

\section{Summary}
\label{chap:conclusion}

In summary, this work investigated different orders for filling SIMD registers with particle values.
An investigation was conducted to observe which vectorization scheme is optimal under which circumstances.
The results show that the optimal behavior and the influencing parameters depend on the MD neighbor identification algorithms and can change during the runtime of the simulation.
In the case of the Verlet Cluster Lists algorithm, only the cluster size was observed to be relevant for identifying the optimal choice of the vectorization pattern.

In the case of the Linked Cells approach, a varying particle distribution over the simulation domain and algorithm-specific parameters, such as the domain traversal mechanism, can change the optimal choice of the interaction order.
Apart from runtime, energy consumption was also considered as a performance metric, and the benchmarks showed interesting differences between both. 
Consequently, the possibility of tuning the SIMD interaction orders at runtime is a valuable extension to achieve performance portability for vectorization.

\section{Acknowledgements}
This work has been supported by the project MaST, which is funded by dtec.bw – Digitalisation and Technology Research Center of the Bundeswehr. dtec.bw is funded by the European Union – NextGenerationEU. The authors also gratefully acknowledge funding for the present work by the Federal Ministry of Education and Research (BMBF, Germany), projects 16ME0610 (WindHPC) and 16ME0653 (3xa). BMBF is funded by the European Union - NextGenerationEU. The computing hardware for the WindHPC project was provided by Intel.
Additionally, the authors thank Ruben Horn for his assistance during the setup of the WindHPC cluster.
Finally, all AutoPas contributors deserve special thanks for their efforts.

%
%
%
%

\end{document}